\documentclass{aa}
\usepackage{elocalloc}
\usepackage{graphicx}
\usepackage{forest}
\useforestlibrary{edges}
\usepackage{xcolor}
\usepackage{listings}
\usepackage{txfonts}
\usepackage{subfigure}
\usepackage{float}
\usepackage{natbib,twoopt}
\usepackage{xspace}
\usepackage{balance}
\usepackage{upgreek}
\usepackage{multirow}
\usepackage{siunitx}
\usepackage{tikz}
\usetikzlibrary{arrows.meta, shapes.geometric, positioning, fit,backgrounds}
\usepackage[normalem]{ulem}
\usepackage{soul}
\usepackage[colorlinks=true, allcolors=blue]{hyperref}
\usepackage{outlines}
\usepackage{mathtools}
\usepackage{multirow}

\usepackage[flushleft]{threeparttable}

\defcitealias{Lesci24}{L24}

\def\apjl{ApJ Lett.}
\def\apjs{ApJ Suppl.}

\def\aap{A\&A}

\bibpunct{(}{)}{;}{a}{}{,}
\DeclareGraphicsExtensions{.eps,.ps,.pdf}

\begin{document}
\title{\texttt{FOMO-Z}. Optimal galaxy classification, photometric redshifts, and reconstruction of population distributions}

\author
{
G. F. Lesci\inst{\ref{2},\ref{1}} 
}

\offprints{G. F. Lesci \\ \email{giorgio.lesci@inaf.it}}

\institute{
  Dipartimento di Fisica e Astronomia ``Augusto Righi'' - Alma Mater Studiorum
  Universit\`{a} di Bologna, via Piero Gobetti 93/2, I-40129 Bologna,
  Italy\label{2}
  \and INAF - Istituto di Radioastronomia, Via Piero Gobetti 101, I-40129, Bologna, Italy\label{1}
}

\date{Received --; accepted --}

\abstract
{}
{
We present \texttt{FOMO-Z} (Flexible Optimizer of Multi-dimensional Operators and photo-$z$s), a set of statistical methods implemented in Python for galaxy classification and probabilistic inference from photometry. The methods combine physically interpretable color selections with uncertainty-based color matching to estimate individual membership probabilities, reconstruct probability density functions of continuous properties, and infer distributions of selected galaxy populations.
}
{
The methods implemented in \texttt{FOMO-Z} use a reference sample with known galaxy properties to perform binary selection and probabilistic inference from photometric colors. They rely on explicit criteria and make no assumptions about the functional form of the color distributions. Binary selections are calibrated using either linear color conditions, particularly suited to background galaxy selection, or flexible color--color grids, seeking to maximize completeness given a threshold on purity. For probabilistic inference, the code identifies reference objects with compatible colors within photometric uncertainties to estimate each test galaxy's probability of satisfying user-defined criteria, as well as probability density functions of continuous properties such as redshift. The code also reconstructs property distributions of selected galaxy populations using photometrically similar galaxies in the reference sample. This framework supports prior weights and reference redshift probability distributions. A built-in feature-selection method identifying informative colors is also introduced.
}
{
As examples of possible applications, we apply \texttt{FOMO-Z} to several astrophysical cases, including background galaxy selection, galaxy population classification, tomographic redshift calibration, and photo-$z$ estimation. The code achieves results consistent with those obtained through other machine-learning approaches, such as self-organizing maps, while remaining computationally efficient, making it suitable for large photometric surveys and future Stage~IV cosmological analyses. The code is publicly available at the link \url{https://github.com/GiorgioLesci/FOMO-Z}.
}
{}


\authorrunning{G. F. Lesci}

\titlerunning{\texttt{FOMO-Z}. Optimal galaxy classification, photometric redshifts, and reconstruction of population distributions}

\maketitle

\section{Introduction}
Current and upcoming imaging surveys, such as those conducted by \textit{Euclid} \citep{Mellier25}, the \textit{Vera C.\ Rubin Observatory} \citep[\textit{Rubin}/LSST;][]{LSST}, and the \textit{Nancy Grace Roman Space Telescope} \citep{Spergel15}, will observe billions of galaxies, for which spectroscopic redshifts (spec-$z$s) will be available for only a small fraction. Photometric redshift (photo-$z$) estimates are therefore essential, but they are affected by significant and potentially non-Gaussian uncertainties. Estimating the underlying redshift distributions of photometrically selected samples is consequently a key requirement for astrophysical and cosmological analyses. More generally, analogous reconstruction problems arise when inferring intrinsic galaxy properties from noisy photometric observables. \\
\indent To address these challenges, a variety of approaches have been developed in recent years. These include machine-learning techniques \citep[e.g.,][]{Wright24,Toni25}, which can efficiently capture complex relations between observables and galaxy properties, as well as statistical approaches based on explicit modeling of the photometric information \citep[e.g.,][]{Bell03,Whitaker11,Medezinski18,Leistedt26}. These approaches differ in their flexibility, interpretability, computational requirements, and dependence on training or reference samples. Explicit selection criteria provide a transparent description of how objects are classified, while methods calibrated on reference samples depend on how well those samples represent the target population. \\
\indent In this paper, we present the open-source Python code \texttt{FOMO-Z} (Flexible Optimizer of Multi-dimensional Operators and photo-$z$s), a machine-learning framework for classifying and selecting objects based on their observable properties. Given an observed \citep[e.g.][]{Weaver22} or simulated \citep[e.g.][]{Viitanen26} reference sample with known properties, \texttt{FOMO-Z} supports both binary selection and probabilistic inference in observable space. It identifies regions preferentially populated by a user-defined target population and estimates individual probabilities of satisfying selection criteria, as well as probability density functions of galaxy properties. The code also reconstructs property distributions of selected test populations using photometrically similar galaxies in the reference sample. The framework supports photometric uncertainties, reference redshift probability distributions, and sample weights.\\
\indent These methods can be applied to independent test samples to select and characterize galaxy populations. Their applications span galaxy evolution \citep{Quai25,Toni25}, cluster weak lensing \citep{Schrabback21,Giocoli21,Ingoglia22,Grandis24,Sereno24}, galaxy-galaxy lensing \citep{Linke20,Heydenreich25,Zhang25}, cosmic shear and galaxy clustering analyses \citep{Heymans21,Wright25_cosmo,Mahony25}. Potential future applications of \texttt{FOMO-Z} include photometric design studies and target selection for spectroscopic or multi-wavelength follow-up observations. The method could be used to evaluate the impact of adding or removing filters on galaxy selection performance, and to optimize selections in scenarios where photo-$z$ information is limited or uncertain. Its flexibility also allows extensions to other astrophysical targets, such as Milky Way stars. \\
\indent The paper is organized as follows. In Sect.~\ref{sec:data}, we present the catalogs used to demonstrate the capabilities and performance of \texttt{FOMO-Z}. In Sect.~\ref{sec:method}, we describe the statistical methods implemented in the code, while Sect.~\ref{sec:results} presents some applications of the code to realistic case studies. Finally, in Sect.~\ref{sec:conclusions}, we summarize our main conclusions. The AB magnitude system is adopted throughout.

\section{Datasets}\label{sec:data}
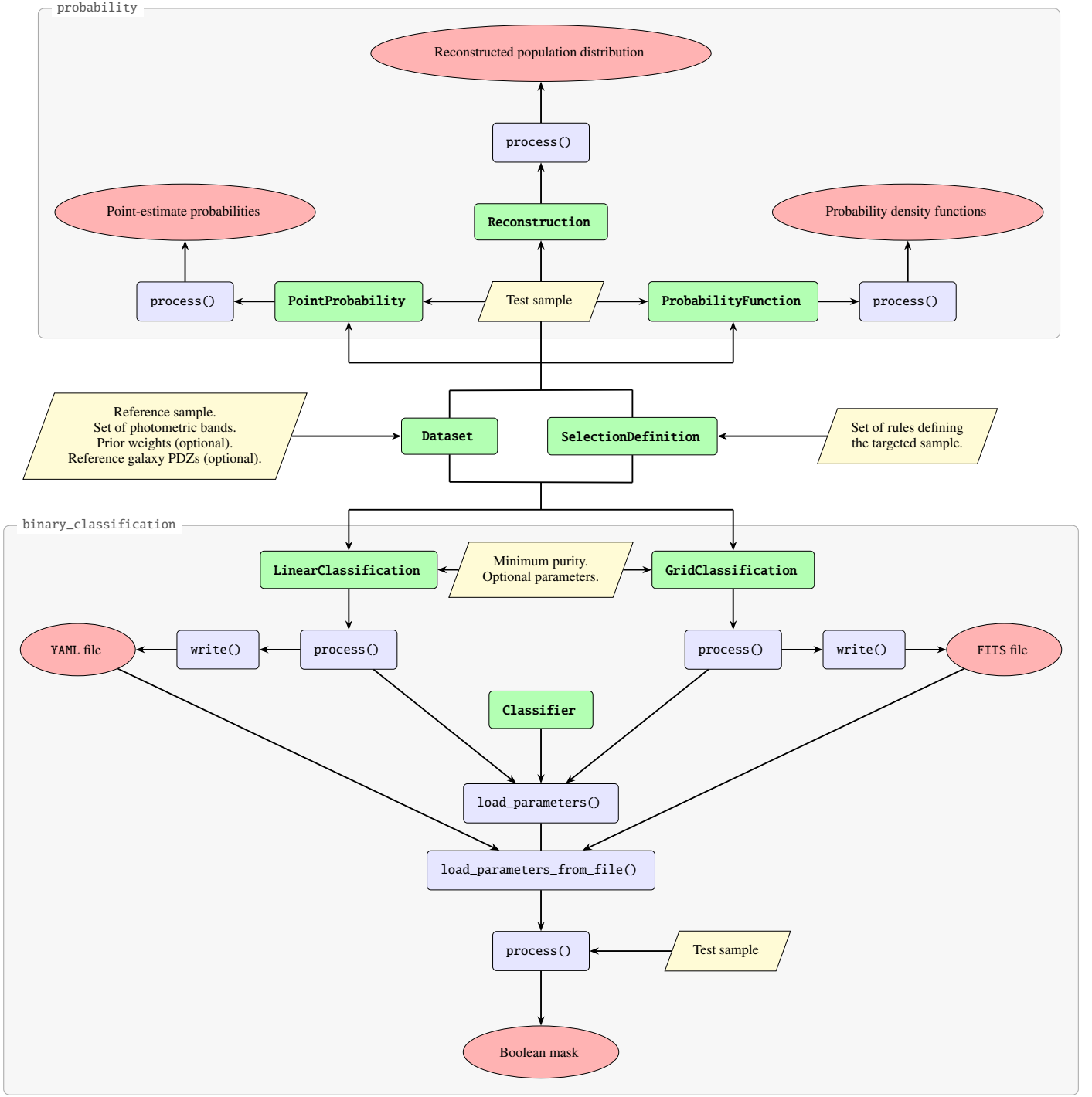
\begin{figure*}[htbp]
\centering
\resizebox{\hsize}{!}{%
\begin{tikzpicture}[
    node distance=3cm and 4cm,
    data/.style={
        trapezium,
        trapezium left angle=70,
        trapezium right angle=110,
        draw,
        minimum width=3.5cm,
        minimum height=1cm,
        align=center,
        font=\Large,
        inner sep=0.5cm,
        fill=yellow!20
    },
    class/.style={
        draw,
        rounded corners,
        minimum width=3.5cm,
        minimum height=1cm,
        align=center,
        font=\Large,
        inner sep=0.5cm,
        fill=green!30
    },
    box/.style={
        draw,
        rounded corners,
        minimum width=1cm,
        minimum height=0.5cm,
        align=center,
        font=\Large,
        inner sep=0.5cm
    },
    method/.style={box, fill=blue!10},
    arrow/.style={-Stealth, line width=1.5pt},
    output/.style={
        ellipse,
        draw,
        minimum width=3.5cm,
        minimum height=1cm,
        align=center,
        font=\Large,
        inner sep=0.5cm,
        fill=red!30
    }
]

\node[data] (inputleft) {
    Reference sample.\\
    Set of photometric bands.\\
    Prior weights (optional).\\
    Reference galaxy PDZs (optional).
};

\node[class, right=of inputleft] (dataset) {
    \textbf{\texttt{Dataset}}
};

\draw[arrow] (inputleft) -- (dataset);

\node[data, right=12cm of dataset] (inputright) {
    Set of rules defining\\
    the targeted sample.
};

\node[class, left=of inputright] (seldef) {
    \textbf{\texttt{SelectionDefinition}}
};

\draw[arrow] (inputright) -- (seldef);

\coordinate (datasetup) at ($(dataset.north)+(0,1cm)$);
\coordinate (seldefup) at ($(seldef.north)+(0,1cm)$);

\draw[line width=1.5pt] (dataset.north) -- (datasetup);
\draw[line width=1.5pt] (seldef.north) -- (seldefup);

\coordinate (midpointup) at ($(datasetup)!0.5!(seldefup)$);

\draw[line width=1.5pt] (datasetup) -- (midpointup);
\draw[line width=1.5pt] (seldefup) -- (midpointup);

\coordinate (mergedup) at ($(midpointup)+(0,1cm)$);
\draw[line width=1.5pt] (midpointup) -- (mergedup);

\coordinate (splitleftup) at ($(mergedup)+(-7cm,0)$);
\coordinate (splitrightup) at ($(mergedup)+(7cm,0)$);

\draw[line width=1.5pt] (mergedup) -- (splitleftup);
\draw[line width=1.5pt] (mergedup) -- (splitrightup);

\node[class, above=1.5cm of splitleftup] (pointprob) {
    \textbf{\texttt{PointProbability}}
};

\node[class, above=1.5cm of splitrightup] (probfunc) {
    \textbf{\texttt{ProbabilityFunction}}
};

\draw[arrow] (splitleftup) -- (pointprob.south);
\draw[arrow] (splitrightup) -- (probfunc.south);

\draw[line width=1.5pt]
    (mergedup) -- ($(pointprob)!0.5!(probfunc)$);

\node[data] (testsampleup) at ($(pointprob)!0.5!(probfunc)$) {
    Test sample
};

\draw[arrow] (testsampleup) -- (pointprob);
\draw[arrow] (testsampleup) -- (probfunc);

\node[class, above=1.5cm of testsampleup] (reconstruction) {
    \textbf{\texttt{Reconstruction}}
};

\draw[arrow]
    (testsampleup.north) -- (reconstruction.south);

\node[method, above=1.5cm of reconstruction] (processrec) {
    \texttt{process()}
};

\draw[arrow]
    (reconstruction.north) -- (processrec.south);

\node[output, above=1.5cm of processrec] (outrec) {
    Reconstructed population distribution
};

\draw[arrow]
    (processrec.north) -- (outrec.south);

\node[method, left=1.5cm of pointprob] (processpoint) {
    \texttt{process()}
};

\node[method, right=1.5cm of probfunc] (processprob) {
    \texttt{process()}
};

\draw[arrow] (pointprob) -- (processpoint);
\draw[arrow] (probfunc) -- (processprob);

\node[
    output,
    above=1.5cm of processpoint,
    minimum height=1.5cm
] (outpoint) {
    Point-estimate probabilities
};

\draw[arrow] (processpoint) -- (outpoint);

\node[
    output,
    above=1.5cm of processprob,
    minimum height=1.5cm
] (outprob) {
    Probability density functions
};

\draw[arrow] (processprob) -- (outprob);

\coordinate (datasetdown) at ($(dataset.south)+(0,-1cm)$);
\coordinate (seldefdown) at ($(seldef.south)+(0,-1cm)$);

\draw[line width=1.5pt] (dataset.south) -- (datasetdown);
\draw[line width=1.5pt] (seldef.south) -- (seldefdown);

\coordinate (midpoint) at ($(datasetdown)!0.5!(seldefdown)$);

\draw[line width=1.5pt] (datasetdown) -- (midpoint);
\draw[line width=1.5pt] (seldefdown) -- (midpoint);

\coordinate (mergeddown) at ($(midpoint)+(0,-1cm)$);
\draw[line width=1.5pt] (midpoint) -- (mergeddown);

\coordinate (splitleft) at ($(mergeddown)+(-7cm,0)$);
\coordinate (splitright) at ($(mergeddown)+(7cm,0)$);

\draw[line width=1.5pt] (mergeddown) -- (splitleft);
\draw[line width=1.5pt] (mergeddown) -- (splitright);

\node[class, below=1.5cm of splitleft] (caliblinear) {
    \textbf{\texttt{LinearClassification}}
};

\node[class, below=1.5cm of splitright] (calibgrid) {
    \textbf{\texttt{GridClassification}}
};

\draw[arrow] (splitleft) -- (caliblinear.north);
\draw[arrow] (splitright) -- (calibgrid.north);

\node[data] (inputcenter) at ($(caliblinear)!0.5!(calibgrid)$) {
    Minimum purity.\\
    Optional parameters.
};

\draw[arrow] (inputcenter) -- (caliblinear);
\draw[arrow] (inputcenter) -- (calibgrid);

\node[method, below=1.5cm of caliblinear] (processlinear) {
    \texttt{process()}
};

\node[method, below=1.5cm of calibgrid] (processgrid) {
    \texttt{process()}
};

\draw[arrow] (caliblinear) -- (processlinear);
\draw[arrow] (calibgrid) -- (processgrid);

\node[method, left=1.5cm of processlinear] (writeleft) {
    \texttt{write()}
};

\node[method, right=1.5cm of processgrid] (writeright) {
    \texttt{write()}
};

\draw[arrow] (processlinear) -- (writeleft);
\draw[arrow] (processgrid) -- (writeright);

\node[output, left=1.5cm of writeleft] (fileleft) {
    \texttt{YAML} file
};

\node[output, right=1.5cm of writeright] (fileright) {
    \texttt{FITS} file
};

\draw[arrow] (writeleft) -- (fileleft);
\draw[arrow] (writeright) -- (fileright);

\coordinate (processmid) at ($(processlinear)!0.5!(processgrid)$);

\node[class, below=1.5cm of processmid] (colorsel) {
    \textbf{\texttt{Classifier}}
};

\node[method, below=2cm of colorsel] (loadparams) {
    \texttt{load\_parameters()}
};

\node[method, below=1cm of loadparams] (loadfromfile) {
    \texttt{load\_parameters\_from\_file()}
};

\draw[arrow] (colorsel) -- (loadparams);
\draw[line width=1.5pt] (loadparams) -- (loadfromfile);

\draw[arrow] (processlinear) -- (loadparams);
\draw[arrow] (processgrid) -- (loadparams);

\draw[arrow] (fileleft) -- (loadfromfile);
\draw[arrow] (fileright) -- (loadfromfile);

\node[method, below=1.5cm of loadfromfile] (processsel) {
    \texttt{process()}
};

\draw[arrow] (loadfromfile) -- (processsel);

\node[data, right=3cm of processsel] (testdata) {
    Test sample
};

\draw[arrow] (testdata) -- (processsel);

\node[
    output,
    below=2cm of processsel,
    minimum height=1.5cm
] (output) {
    Boolean mask
};

\draw[arrow] (processsel) -- (output);

\begin{scope}[on background layer]
    \node[
        draw=black!35,
        fill=black!3,
        line width=0.7pt,
        rounded corners=6pt,
        inner sep=0.6cm,
        fit=(pointprob)(probfunc)(testsampleup)
            (processpoint)(processprob)(outpoint)(outprob)
            (reconstruction)(processrec)(outrec)
    ] (probabilitygroup) {};

    \node[
        draw=black!35,
        fill=black!3,
        line width=0.7pt,
        rounded corners=6pt,
        inner sep=0.6cm,
        fit=(caliblinear)(calibgrid)(inputcenter)
            (processlinear)(processgrid)
            (writeleft)(writeright)(fileleft)(fileright)
            (colorsel)(loadparams)(loadfromfile)
            (processsel)(testdata)(output)
    ] (selectiongroup) {};
\end{scope}

\node[
    anchor=west,
    fill=white,
    inner xsep=0.2cm,
    inner ysep=0.1cm,
    font=\Large,
    text=black!65
] at ([xshift=0.5cm]probabilitygroup.north west) {
    \texttt{probability}
};

\node[
    anchor=west,
    fill=white,
    inner xsep=0.2cm,
    inner ysep=0.1cm,
    font=\Large,
    text=black!65
] at ([xshift=0.5cm]selectiongroup.north west) {
    \texttt{binary\_classification}
};

\end{tikzpicture}%
}
\caption{
Overview of the \texttt{FOMO-Z} workflows for probabilistic inference
and binary selection, grouped by subpackage.
Input data are shown in yellow trapezoids, classes in green rectangles,
methods in blue rectangles, and outputs in red ellipses.
Arrows indicate data flow and method dependencies.
The shared \texttt{Dataset} and \texttt{SelectionDefinition} objects
provide the reference sample and target selection criteria.
The \texttt{probability} subpackage estimates individual selection
probabilities through \texttt{PointProbability} and probability density
functions through \texttt{ProbabilityFunction}.
Population distributions for selected test samples are reconstructed
through \texttt{Reconstruction}. Only the main processing methods are shown.
The \texttt{binary\_classification} subpackage calibrates binary selections
through \texttt{LinearClassification} and \texttt{GridClassification}.
Calibration parameters can be passed directly to \texttt{Classifier}
or saved to file and subsequently loaded.
Applying the calibrated selection to a test sample produces a Boolean mask.
}
\label{fig:flowchart}
\end{figure*}
To test the code implementations described in this paper, we used the four different public catalogs presented in the following. These datasets were adopted as reference samples for the science cases presented in Sect.~\ref{sec:results}. \\
\indent The COSMOS \texttt{CLASSIC} catalog from \citet[][COSMOS2020]{Weaver22} was used as the reference sample to validate the selections of background galaxies (Sect.~\ref{sec:results:background}). This catalog achieves a photometric redshift precision of $\sigma_z/(1+z) \simeq 0.01$ for sources with $i < 22.5$. We adopted galaxy redshift point estimates corresponding to the median of the posterior redshift probability distribution derived with the \texttt{LePhare} code \citep{Arnouts02,Ilbert06}. To correct for Galactic foreground extinction, all photometric bands are de-reddened prior to the analysis. For each object, we used the Milky Way color excess $E(B-V)$ provided by the COSMOS2020 catalog and assumed a total-to-selective extinction ratio $R_V = 3.1$. The extinction in each photometric band was then computed at the filter effective wavelength using the reddening law of \citet{FM07}. We considered Kron aperture magnitudes \citep{Kron1980}, selecting galaxies with photo-$z$s derived from at least 30 bands and having $i < 25$, where the $i$ magnitude is obtained from the Hyper Suprime-Cam \citep[HSC,][]{HSC}. This selection ensures a photo-$z$ precision better than $\sigma_z/(1+z)=0.025$ and yields a sample of about 200\,000 galaxies. \\
\indent To select quiescent and star-forming galaxies (Sect.\ \ref{sec:results:quiescent}), we used the COSMOS sample of \citet[][COSMOS15]{Laigle16}. Applying the galaxy selections and magnitude corrections described in \citet{Toni25} yields a sample of approximately 300\,000 galaxies. We then applied the \texttt{FOMO-Z} selection calibrated on this sample to the COSMOS-Web sample of \citet{Toni25}, which contains 388\,173 galaxies and is based on the COSMOS-Web catalog of \citet{Shuntov26}. The latter combines previous COSMOS data releases with \textit{JWST}/NIRCam imaging in the F115W, F150W, F277W, and F444W filters \citep{Rieke23,Franco26}, reaching $5\sigma$ point-source depths of 27.5--28.2 mag and covering an area of 0.54 deg$^2$. \\
\indent To compare \texttt{FOMO-Z} with self-organizing maps (Sect.~\ref{sec:results:soms}) and to maximize the signal-to-noise (S/N) of weak lensing profiles (Sect.\ \ref{sec:results:maximize}), we used the spectroscopic compilation adopted to reconstruct galaxy redshift distributions in the fifth data release of the Kilo-Degree Survey (KiDS-Legacy; \citealt{Wright25}). This compilation, referred to as KiDZ\footnote{\url{https://kids.strw.leidenuniv.nl/DR5/legacy_kidz.php}} \citep{vandenBusch22,Wright24}, comprises approximately 120\,000 galaxies with spectroscopic redshifts drawn from 22 surveys. The sample spans an effective area of 19.3~deg$^2$ and is matched to KiDS photometry, which comprises $u$, $g$, $r$, and $i$ bands from the OmegaCAM wide-field camera \citep{OmegaCAM} mounted on the VLT Survey Telescope \citep{VST}, as well as $z$, $Y$, $J$, $H$, and $K_{\rm s}$ bands from the VISTA Kilo degree INfrared Galaxy (VIKING) survey \citep{VIKINGS,VISTA}. Photometric redshifts are also available for each galaxy in the catalog.

\section{Methods}\label{sec:method}
The methods presented in this work are designed to classify objects and characterize samples directly from observable quantities. Given a reference sample with known physical properties, \texttt{FOMO-Z} supports both binary selection and probabilistic inference for objects in independent samples. Binary selections identify regions of observable space preferentially populated by the target population, while probabilistic methods estimate individual probabilities of satisfying selection criteria and reconstruct probability density functions of galaxy properties. Population distributions can also be reconstructed for selected test samples using photometrically compatible reference objects. In this work, we use galaxy colors as inputs and focus on redshift and star-formation activity as target properties. \\
\indent As described in the following sections, binary selection methods are implemented in the \texttt{binary\_classification} subpackage, while probabilistic inference and population reconstruction are implemented in \texttt{probability}. The \texttt{data} subpackage handles input datasets and the definition of selection criteria. A flowchart illustrating the structure of the code is shown in Fig.~\ref{fig:flowchart}. Required Python packages, installation guidelines, and example Python scripts are provided in Appendix~\ref{app:usage}, while code performance and memory usage are detailed in Appendix~\ref{app:performance}.

\subsection{\texttt{\textup{Dataset}}: Input data handling}\label{sec:method:dataset}
The \texttt{Dataset} class of the \texttt{data} subpackage represents reference and test galaxy samples. It stores photometric magnitudes and any additional properties required for classification or reconstruction. The class standardizes the input data, handles missing values, and constructs colors from the supplied photometric bands. Magnitude uncertainties are used by the probabilistic methods to determine whether reference and test objects have compatible colors. Given an input set of photometric bands, the \texttt{Dataset} class constructs all distinct pairs of photometric bands, retaining one color per pair, whose number is given by
\begin{equation}\label{eq:N_color}
N_{\rm color} = \frac{N_{\rm band}!}{(N_{\rm band}-2)!\,2!},
\end{equation}
where $N_{\rm band}$ is the number of photometric bands provided by the user. Alternatively, the user may supply an explicit list of colors, allowing the analysis to focus on specific color combinations. \\
\indent As described below, binary classification methods calibrate selection regions in two-dimensional color--color spaces, whereas probabilistic methods identify compatible objects by comparing their individual colors. Given the set of available colors, the number of distinct color--color spaces considered in binary selection calibration is
\begin{equation}\label{eq:N_space}
N_{\rm space} = \frac{N_{\rm color}!}{(N_{\rm color}-2)!\,2!}\,.
\end{equation}
Object weights can be supplied for both reference and test samples. Reference prior weights, denoted as $w_{\rm prior}$, determine the relative contributions of reference objects. This can account, for example, for spectroscopic redshift selection functions, spatial variations in survey depth, or, in weak lensing analyses, the relative significance of shape measurements \citep[see, e.g.,][]{Wright25}. In population reconstruction, test-sample weights determine each object's contribution to the target distribution, as described in Sect.~\ref{sec:method:reconstruction}. Galaxy redshift probability density functions (PDZs) may also be supplied. For reference samples, their integrals over specified intervals provide redshift-selection weights (Sect.~\ref{sec:method:sel}) or contributions to reconstructed redshift bins (Sect.~\ref{sec:method:reconstruction}). The use of PDZs accounts for reference redshift uncertainties rather than relying solely on point estimates.

\subsection{\texttt{\textup{SelectionDefinition}}: Definition of selection criteria}\label{sec:method:sel}
The classification and reconstruction classes rely on \texttt{SelectionDefinition} to specify the criteria defining the target sample. Selection rules may involve any numerical or categorical property available in the reference sample, including an optional redshift constraint, denoted as $\mathcal{R}_z$:
\begin{equation}\label{eq:zl}
\mathcal{R}_z : z_{\rm g} \in (z_{\rm l}, z_{\rm u})\,,
\end{equation}
where $z_{\rm g}$ is the galaxy redshift, while $z_{\rm l}$ and $z_{\rm u}$ are the lower and upper redshift limits, respectively. Additional selection rules, collectively denoted as $\mathcal{R}_{\rm add}$, may be defined and combined with $\mathcal{R}_z$ through a logical `and', denoted as $\land$. The full rule set is therefore
\begin{equation}\label{eq:R}
\mathcal{R} : \mathcal{R}_z\land\mathcal{R}_{\rm add}\,.
\end{equation}
The rules in $\mathcal{R}_{\rm add}$ may include, for example, galaxy stellar mass (e.g., $M^* > 10^{12}\,{\rm M}_\odot$), magnitude limits (e.g., $u < 23$), or object classifications (e.g., requiring the presence of an X-ray counterpart). The binary selection and probabilistic methods described in the following sections use the reference sample to relate galaxy colors to the properties specified by $\mathcal{R}$, enabling both the selection of target objects and the estimation of individual probabilities of satisfying these criteria. \\
\indent The rule set $\mathcal{R}$ assigns a selection weight to each galaxy in the reference sample:
\begin{equation}\label{eq:w_selection}
w_{\rm sel}^j(\mathcal{R}) = w_{\rm prior}^j\,w^j(\mathcal{R}_z)\,w^j(\mathcal{R}_{\rm add})\,,
\end{equation}
where $j$ indexes reference objects and $w_{\rm prior}^j$ is the prior weight (Sect.~\ref{sec:method:dataset}), which equals 1 by default in the \texttt{FOMO-Z} code. If no redshift constraint is imposed, $w^j(\mathcal{R}_z)=1$, and the selection is determined entirely by the additional rules. Similarly, $w^j(\mathcal{R}_{\rm add})=1$ when no additional rules are specified. In Eq.\ \eqref{eq:w_selection}, the contribution from the redshift rule, denoted as $w^j(\mathcal{R}_z)$, takes the following form:
\begin{equation}\label{eq:R_z}
w^j(\mathcal{R}_z) = \int_{z_{\rm l}}^{z_{\rm u}} \mathrm{d}z_{\rm g}\, p^j(z_{\rm g})\,.
\end{equation}
Here, $p^j(z_{\rm g})$ is the reference galaxy's PDZ, normalized over the supplied redshift grid (Sect.~\ref{sec:method:dataset}). When a redshift constraint is imposed but reference PDZs are unavailable, membership is determined from the redshift point estimate:
\begin{equation}
w^j(\mathcal{R}_z) = \delta^j(\mathcal{R}_z)\,,
\end{equation}
where $\delta^j(\mathcal{R}_z)=1$ if the galaxy's redshift point estimate satisfies Eq.\ \eqref{eq:zl}, and $0$ otherwise. For the additional rules in Eq.\ \eqref{eq:w_selection}, we define
\begin{equation}\label{eq:R_add}
w^j(\mathcal{R}_{\rm add}) = \prod_k \delta^j(\mathcal{R}_{\rm add}^k)\,,
\end{equation}
where $\mathcal{R}_{\rm add}^k$ is the $k$th rule, while $\delta^j(\mathcal{R}_{\rm add}^k) = 1$ if the $j$th reference galaxy satisfies $\mathcal{R}_{\rm add}^k$ and 0 otherwise. For population reconstruction, the initial selection instead defines an observed interval applied to both reference and test samples. This distinction is described in Sect.~\ref{sec:method:reconstruction}.

\subsection{\texttt{\textup{LinearClassification}}: Linear color cuts}\label{sec:method:linear}
\begin{figure*}[t]
\centering\includegraphics[width = \hsize-1.8cm, height = 9cm] {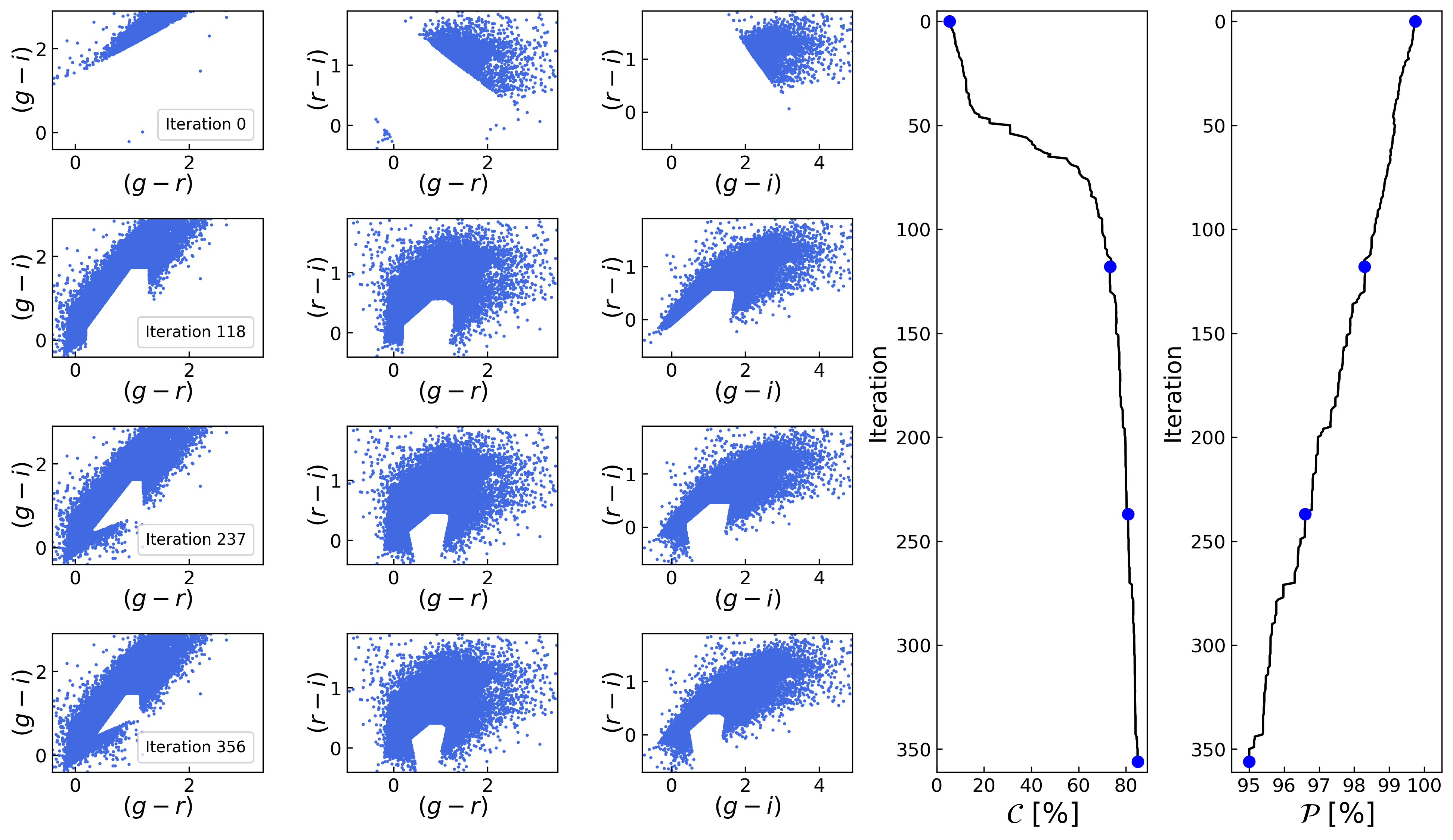}
\caption{Visualization of the iterative process implemented in the \texttt{LinearClassification} class (Sect.\ \ref{sec:method:linear}). Here, we considered $gri$ photometry and assumed a limiting lower redshift of $z_{\rm l}=0.3$, a global purity threshold of $\mathcal{P}_{\rm min}=0.95$, and a step size of $\delta\mathcal{P}_{\rm step}=10^{-4}$. \textit{Left}: $(g-i)$ -- $(g-r)$ (left), $(r-i)$ -- $(g-r)$ (middle), and $(r-i)$ -- $(g-i)$ (right) color--color spaces. The blue dots represent the galaxies selected in the reference sample. The iteration number, reported in the legends, increases from top to bottom. \textit{Right}: iteration number as a function of the global completeness (left) and purity (right). The blue dots are in correspondence of the iterations shown in the left panels.}
\label{fig:iterations}
\end{figure*}
\begin{figure}[t!]
\centering\includegraphics[width = \hsize-0.8cm, height = 8cm] {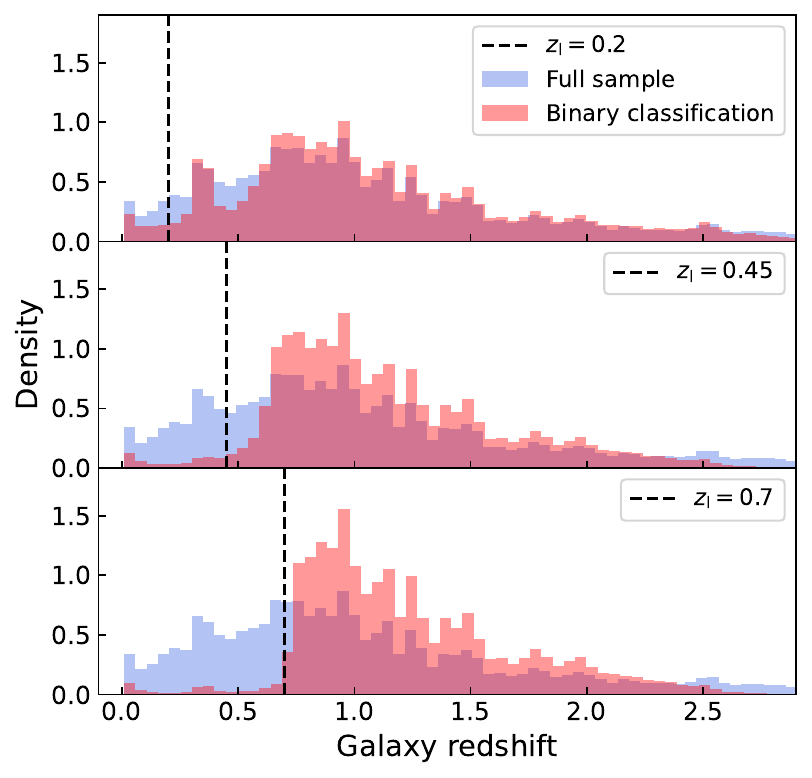}
\caption{Redshift distributions of the COSMOS20 reference sample before (blue) and after (red) applying the binary color selections calibrated with \texttt{LinearClassification}, using a minimum purity threshold of $\mathcal{P}_{\rm min}=0.97$. The selections target galaxies at $z>z_{\rm l}$, with $z_{\rm l}=0.2$ (top), $0.45$ (middle), and $0.7$ (bottom). Black vertical dashed lines mark the corresponding redshift thresholds.}
\label{fig:nz}
\end{figure}

One of the binary classification methods included in the \texttt{binary\_classification} subpackage, implemented in the \texttt{LinearClassification} class, is based on linear relations in color--color spaces. This method is specifically designed for the selection of galaxies that occupy the outskirts of color--color distributions. For instance, it is particularly useful in the selection of background galaxies for weak lensing measurements, where foreground galaxies must be excluded to avoid diluting the lensing signal. An alternative binary classification algorithm, suitable for other science cases, is described in Sect.\ \ref{sec:method:regions}. The methodological framework implemented in \texttt{LinearClassification} and detailed in this section was also presented by \citet[][hereafter L24]{Lesci24}. Improvements over the methods presented in \citetalias{Lesci24} are discussed in Appendix \ref{app:L24}. \\
\indent Given the photometric reference sample provided in input to the \texttt{Dataset} class (Sect.\ \ref{sec:method:dataset}), the methods of the \texttt{LinearClassification} class consider the following conditions in each color--color space:
\begin{align}\label{eq:all_cond}
&x > c_1\,\lor\,\nonumber\\
&x < c_2\,\lor\,\nonumber\\
&y > c_3\,\lor\,\nonumber\\
&y < c_4\,\lor\,\nonumber\\
&x > s_1 y + c_5\,\lor\,\nonumber\\
&x > s_2 y + c_6\,\lor\,\nonumber\\
&x < s_3 y + c_7\,\lor\,\nonumber\\
&x < s_4 y + c_8,
\end{align}
where $x$ and $y$ are two different colors, $\lor$ is the logical `or' operator, while $c_p$ and $s_q$ are free parameters. Equation \eqref{eq:N_space} defines the total number of color–-color spaces over which Eq.\ \eqref{eq:all_cond} is evaluated. Nominally, we impose $c_1,\ldots,c_8\in(-\infty,+\infty)$, $s_1$ and $s_3\in(0,+\infty)$, $s_2$ and $s_4\in(-\infty,0)$. In practice, the domain of the $c_1,\ldots,c_8$ parameters is efficiently sampled employing the data-driven approach described in Appendix \ref{app:sampling}, while the $s_q$ samples are directly set by the user. \\
\indent We remark that all conditions in Eq.\ \eqref{eq:all_cond} are treated as independent, and that their combination defines an irregular octagon in each $(x,y)$ space, which potentially contains contaminants. However, since some of the conditions in Eq. \eqref{eq:all_cond} may not meet purity or completeness requirements, only a subset may be considered for the final color selection, as detailed in the following.\\
\indent For any color pair $(x,y)$, each condition $k$ defined in
Eq.\ \eqref{eq:all_cond}, and each reference galaxy $j$, the \texttt{process} method of the \texttt{LinearClassification} class computes the selection binary mask $\mathcal{S}_{xy,k}^j(\vec{p}_{xy,k})$, defined as follows
\begin{equation}\label{eq:S_mask}
\mathcal{S}_{xy,k}^j(\vec{p}_{xy,k}) = 
\begin{cases}
1 & \text{if galaxy $j$ meets the $k$th condition in $(x,y)$} \\
0 & \text{otherwise}
\end{cases}.
\end{equation}
Here, $\vec{p}_{xy,k}$
is the set of parameters in the $k$th color condition (i.e.\ the $c_p$ and $s_q$
values). Then, the true positives take the form
\begin{equation}\label{eq:TP}
\mathrm{TP}_{xy,k}(\mathcal{R},\vec{p}_{xy,k}) = \sum_j w_{\rm sel}^j(\mathcal{R}) \, \mathcal{S}_{xy,k}^j(\vec{p}_{xy,k}) \,,
\end{equation}
where $\mathcal{R}$ is the set of rules defining the targeted sample (Eq.\ \ref{eq:R}), which defines the weight $w_{\rm sel}^j$ (Eq.\ \ref{eq:w_selection}), while $j$ runs over the galaxies in the reference sample. Analogous definitions hold for the false positives,
\begin{equation}\label{eq:FP}
\mathrm{FP}_{xy,k}(\mathcal{R},\vec{p}_{xy,k}) = \left[\sum_j w_{\rm prior}^j \, \mathcal{S}_{xy,k}^j(\vec{p}_{xy,k})\right] - \mathrm{TP}_{xy,k}(\mathcal{R},\vec{p}_{xy,k}) \,,
\end{equation}
the false negatives,
\begin{equation}\label{eq:FN}
\mathrm{FN}_{xy,k}(\mathcal{R},\vec{p}_{xy,k}) = \sum_j w_{\rm sel}^j(\mathcal{R}) \, \left[1-\mathcal{S}_{xy,k}^j(\vec{p}_{xy,k})\right] \,,
\end{equation}
and the true negatives,
\begin{equation}\label{eq:TN}
\mathrm{TN}_{xy,k}(\mathcal{R},\vec{p}_{xy,k}) = \left[\sum_j w_{\rm prior}^j \, [1-\mathcal{S}_{xy,k}^j(\vec{p}_{xy,k})]\right] - \mathrm{FN}_{xy,k}(\mathcal{R},\vec{p}_{xy,k}) \,.
\end{equation}
In Eqs.\ \eqref{eq:FP} and \eqref{eq:TN}, $w_{\rm prior}^j$ is the prior weight of the $j$th galaxy (Sect.\ \ref{sec:method:sel}). From Eqs.\ \eqref{eq:TP}--\eqref{eq:TN}, we define the completeness of the $k$th condition in Eq.\ \eqref{eq:all_cond} as
\begin{equation}\label{eq:C}
\mathcal{C}_{xy,k}(\mathcal{R},\vec{p}_{xy,k}) = \frac{\mathrm{TP}_{xy,k}(\mathcal{R},\vec{p}_{xy,k})} {\mathrm{TP}_{xy,k}(\mathcal{R},\vec{p}_{xy,k}) + \mathrm{FN}_{xy,k}(\mathcal{R},\vec{p}_{xy,k})}\,,
\end{equation}
as well as the purity,
\begin{equation}\label{eq:P}
\mathcal{P}_{xy,k}(\mathcal{R},\vec{p}_{xy,k}) = \frac{\mathrm{TP}_{xy,k}(\mathcal{R},\vec{p}_{xy,k})} {\mathrm{TP}_{xy,k}(\mathcal{R},\vec{p}_{xy,k}) + \mathrm{FP}_{xy,k}(\mathcal{R},\vec{p}_{xy,k})}\,,
\end{equation}
and the failure rate,
\begin{equation}\label{eq:F}
\mathcal{F}_{xy,k}(\mathcal{R},\vec{p}_{xy,k}) = \frac{\mathrm{FP}_{xy,k}(\mathcal{R},\vec{p}_{xy,k})} {\mathrm{FP}_{xy,k}(\mathcal{R},\vec{p}_{xy,k}) + \mathrm{TN}_{xy,k}(\mathcal{R},\vec{p}_{xy,k})}\,.
\end{equation}
In this weighted formulation, all quantities are expressed as ratios of weighted counts rather than discrete counts. Completeness represents the fraction of objects satisfying $\mathcal{R}$ that are recovered, while purity is the fraction of selected objects that satisfy $\mathcal{R}$. The failure rate (i.e., the false positive rate) is the fraction of selected non-targets. \\
\indent After the evaluation of Eqs.\ \eqref{eq:C}--\eqref{eq:F} for any $(x,y)$ pairs, color condition indices $k$, and parameter sets $\vec{p}_{xy,k}$, the \texttt{LinearClassification} class identifies the optimal combination of color conditions that maximizes the global completeness at fixed global purity. In particular, global completeness, purity, and failure rate are defined as follows:
\begin{equation}\label{eq:C_glob}
\mathcal{C}(\mathcal{R},\vec{P}) = \frac{\mathrm{TP}(\mathcal{R},\vec{P})} {\mathrm{TP}(\mathcal{R},\vec{P}) + \mathrm{FN}(\mathcal{R},\vec{P})}\,,
\end{equation}
\begin{equation}\label{eq:P_glob}
\mathcal{P}(\mathcal{R},\vec{P}) = \frac{\mathrm{TP}(\mathcal{R},\vec{P})} {\mathrm{TP}(\mathcal{R},\vec{P}) + \mathrm{FP}(\mathcal{R},\vec{P})}\,,
\end{equation}
\begin{equation}\label{eq:F_glob}
\mathcal{F}(\mathcal{R},\vec{P}) = \frac{\mathrm{FP}(\mathcal{R},\vec{P})} {\mathrm{FP}(\mathcal{R},\vec{P}) + \mathrm{TN}(\mathcal{R},\vec{P})}\,.
\end{equation}
Here, $\vec{P}$ is the ensemble of $\vec{p}_{xy,k}$ sets relative to each color condition $k$ (Eq.\ \ref{eq:all_cond}) across all the $N_{\rm space}$ (Eq.\ \ref{eq:N_space}) $(x,y)$ spaces. The quantities $\mathrm{TP}$, $\mathrm{FN}$, $\mathrm{FP}$, and $\mathrm{TN}$ are obtained by substituting $\mathcal{S}_{xy,k}^j(\vec{p}_{xy,k})$ in Eqs.\ \eqref{eq:TP}--\eqref{eq:TN} with $\mathcal{S}^j(\vec{P})$, which equals 1 if the $j$th galaxy satisfies at least one color condition, and 0 otherwise. \\
\indent The optimal combination of color conditions is determined through the iterative procedure described below. The following parameters are set by the user: a global purity threshold, $\mathcal{P}_{\rm min}$, a step size $\delta\mathcal{P}_{\rm step}$, and a minimum completeness threshold imposed to each color condition, $\mathcal{C}_{\rm min}$. By default, \texttt{FOMO-Z} assumes $\mathcal{C}_{\rm min} = 5\%$. Enforcing that each color condition yields a completeness greater than $\mathcal{C}_{\rm min}$ suppresses random fluctuations and reduces the statistical uncertainties in selection performance metrics. Given a set of selection criteria $\mathcal{R}$, the optimal color conditions are identified through the following procedure, implemented in the \texttt{process} method of the \texttt{LinearClassification} class.
\begin{enumerate}
    \item Update the iterative purity threshold $\mathcal{P}_{\rm iter}^n=1 - n\,\delta\mathcal{P}_{\rm step}$, where $n$ is the iteration index, starting from 0.
    \item For any $(x,y)$ color pair, select the $k$th condition (see Eq.\ \ref{eq:all_cond}) if there exists at least one parameter set $\vec{p}_{xy,k}$ that yields $\mathcal{P}_{xy,k}(\mathcal{R},\vec{p}_{xy,k})>\mathcal{P}_{\rm iter}^n$ and $\mathcal{C}_{xy,k}(\mathcal{R},\vec{p}_{xy,k})>\mathcal{C}_{\rm min}$. Cases where $\mathcal{F}_{xy,k}(\mathcal{R},\vec{p}_{xy,k})=1$ are excluded, as they do not represent actual selections.
    \item For each color condition selected on step 2, define the optimal $\vec{p}_{xy,k}$ parameter set as the one that maximizes $\mathcal{C}_{xy,k}(\mathcal{R},\vec{p}_{xy,k})$.
    \item Combine, through the $\lor$ operator, the selected conditions in all $(x,y)$ spaces which have optimal $\vec{p}_{xy,k}$ parameters. This yields the global completeness, $\mathcal{C}$ (Eq.\ \ref{eq:C_glob}), purity, $\mathcal{P}$ (Eq.\ \ref{eq:P_glob}), and failure rate, $\mathcal{F}$ (Eq.\ \ref{eq:F_glob}).
    \item Repeat from point 1 as long as $\mathcal{P}>\mathcal{P}_{\rm min}$.
\end{enumerate}
Figure \ref{fig:iterations} shows an example of this optimization algorithm. The \texttt{process} method also computes bootstrap errors for $\mathcal{C}$, $\mathcal{P}$, and $\mathcal{F}$ by resampling the reference sample and applying the optimized selection to each bootstrap realization. We note that these bootstrap error estimates do not account for sample variance. Quantifying this contribution would require multiple independent reference samples.\\
\indent The calibration process described in this section is designed for background galaxy selections. Multiple redshift lower limits $z_{\rm l}$ (see Eq.\ \ref{eq:zl}) may be provided as input to the \texttt{SelectionDefinition} class (Sect.\ \ref{sec:method:sel}), and a distinct minimum purity threshold $\mathcal{P}_{\rm min}$ can be assigned to each of them. Figure~\ref{fig:nz} shows the redshift distributions of the COSMOS20 reference sample before and after applying the binary color selections calibrated with \texttt{LinearClassification}, for three example values of the lower redshift threshold $z_{\rm l}$. If multiple $z_{\rm l}$ points are provided, the parameters $c_p$ and $s_q$ of the optimal color conditions are interpolated as a function of $z_{\rm l}$. See Appendix \ref{app:interp} for details.\\
\indent As shown in Fig.\ \ref{fig:flowchart}, the output of this process, that is a set of color conditions with optimal parameters, can be written to a \texttt{YAML} file. This output can be passed to the \texttt{Classifier} class to apply the calibrated selection to a test dataset. The \texttt{process} method of \texttt{Classifier} accepts a photometric test sample and a set of redshift intervals as input. Because the $c_p$ and $s_q$ parameters of the optimal linear color cuts can be interpolated as functions of the redshift lower limit $z_{\rm l}$, users can select galaxies using any $z_{\rm l}$ value within the range originally provided to the \texttt{LinearClassification} class. The \texttt{process} method of \texttt{Classifier} generates a list of boolean arrays, each corresponding to a specific redshift range and indicating the objects that satisfy the calibrated selection criteria. This method enables users to efficiently apply photometric selections to their data.

\subsection{\texttt{\textup{GridClassification}}: Grid-based color selections}\label{sec:method:regions}
\begin{figure}[t]
\centering\includegraphics[width = \hsize-1.5cm, height = \hsize -2.5cm] {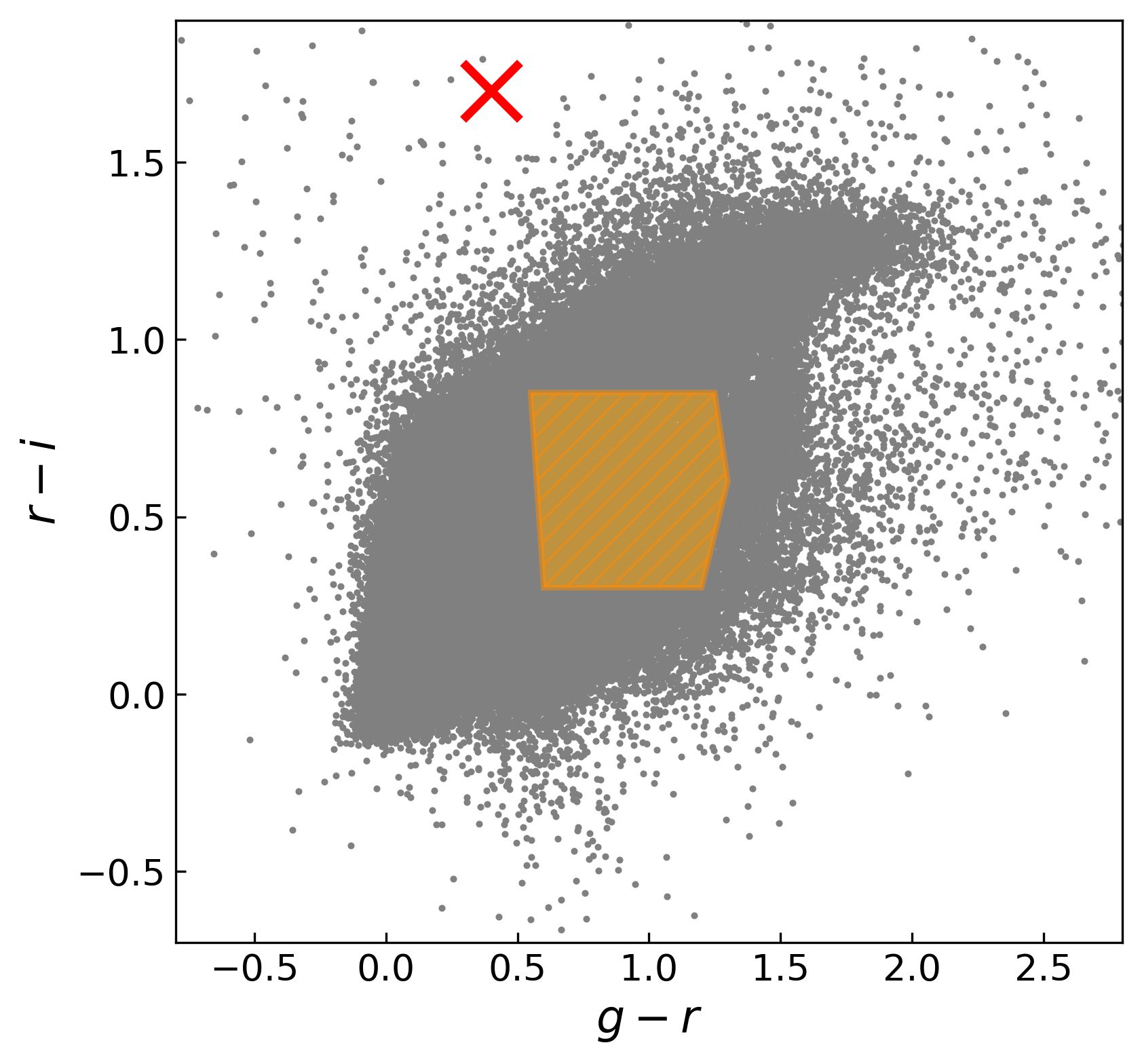}
\caption{Visualization of a conceptual difference between linear color cuts (Sect.\ \ref{sec:method:linear}) and the grid-based selections (Sect.\ \ref{sec:method:regions}) implemented in the \texttt{binary\_classification} subpackage. Gray points represent reference galaxies. A galaxy with colors corresponding to the red cross cannot be selected through grid-based selections, due to poor statistics in that region of the color--color space. Galaxies in such regions can be selected via linear color cuts. In contrast, linear selections cannot define a selection region corresponding to the orange hatched patch, because linear color conditions are connected through the logical `or' operator. In these cases, grid-based binary selections are better suited.}
\label{fig:conceptual_diff}
\end{figure}

In addition to the linear color selection method described in the previous section, the \texttt{binary\_classification} subpackage provides an alternative binary classification algorithm implemented in the \texttt{GridClassification} class, which identifies optimal regions in color-–color spaces by directly exploring discretized grids. Domains and precisions of such grids are detailed in Appendix \ref{app:sampling}. Given a grid in the color--color space $(x,y)$, the code computes the following probability of finding objects with properties defined by $\mathcal{R}$ (analogous to the purity definition in Eq.\ \ref{eq:P}):
\begin{equation}\label{eq:p_selection}
p_{\rm sel}(\mathcal{R},\delta x,\delta y) = \frac{\sum_j w_{\rm sel}^j(\mathcal{R})\,\mathcal{S}^j(\delta x,\delta y)}{\sum_j w_{\rm prior}^j\,\mathcal{S}^j(\delta x,\delta y)} \,.
\end{equation}
Here, $\delta x$ and $\delta y$ represent bins of $x$ and $y$ colors, respectively. The index $j$ runs over the number of reference galaxies, while $ w_{\rm prior}^j$ and $ w_{\rm sel}^j(\mathcal{R})$ are the prior and selection weights, respectively (Eq.\ \ref{eq:w_selection}). Furthermore, $\mathcal{S}^j(\delta x,\delta y)=1$ if the $j$th galaxy falls in the $(\delta x,\delta y)$ grid bin, and 0 otherwise. This definition is analogous to the selection binary mask used for linear color cuts (Eq.\ \ref{eq:S_mask}).\\
\indent Equation \eqref{eq:p_selection} is computed in $N_{\rm space}$ color--color spaces (Eq.\ \ref{eq:N_space}) and leads to the following mask definition, $\mathcal{M}_{\rm sel}$:
\begin{equation}\label{eq:M_selection}
\mathcal{M}_{\rm sel}(\mathcal{R},\delta x, \delta y,p_{\rm min}) = 
\begin{cases}
1 & \text{if $p_{\rm sel}(\mathcal{R},\delta x, \delta y) > p_{\rm min}$}\\
0 & \text{otherwise}
\end{cases}\,.
\end{equation}
Here, $p_{\rm min}$ is a threshold applied to the probability $p_{\rm sel}$, and is a key quantity in the iterative process described below. From Eq.\ \eqref{eq:M_selection}, we define the galaxy selection mask, $\mathcal{S}^j_{\rm sel}$, as
\begin{equation}\label{eq:S_selection}
\mathcal{S}^j_{\rm sel}(\mathcal{R},p_{\rm min}) = 
\begin{cases}
\multirow{2}{*}{$1$} & \text{if galaxy $j$ falls in a grid bin where} \\
                     & \text{$\mathcal{M}_{\rm sel}(\mathcal{R},\delta x,\delta y,p_{\rm min}) = 1$} \\[1ex]
0                    & \text{otherwise} \\[1ex]
\end{cases}\,.
\end{equation}
In words, a galaxy is selected through Eq.\ \eqref{eq:S_selection} if its associated probability of satisfying $\mathcal{R}$ is larger than $p_{\rm min}$ in at least one color--color space. \\
\indent From Eq.\ \eqref{eq:S_selection}, we can express true positives, TP, false positives, FP, false negatives, FN, and true negatives, TN, as follows:
\begin{align}
\mathrm{TP}(\mathcal{R},p_{\rm min}) &= \sum_j w^j_{\rm sel}(\mathcal{R})\,\mathcal{S}^j_{\rm sel}(\mathcal{R},p_{\rm min})\,,\\
\mathrm{FP}(\mathcal{R},p_{\rm min}) &= \left[\sum_j w^j_{\rm prior}\,\mathcal{S}^j_{\rm sel}(\mathcal{R},p_{\rm min})\right] - \mathrm{TP}(\mathcal{R},p_{\rm min})\,,\\
\mathrm{FN}(\mathcal{R},p_{\rm min}) & = \sum_j w_{\rm sel}^j(\mathcal{R}) \, \left[1-\mathcal{S}^j_{\rm sel}(\mathcal{R},p_{\rm min})\right] \,,\\
\mathrm{TN}(\mathcal{R},p_{\rm min}) & = \left[\sum_j w_{\rm prior}^j \, [1-\mathcal{S}^j_{\rm sel}(\mathcal{R},p_{\rm min})]\right] - \mathrm{FN}(\mathcal{R},p_{\rm min}) \,.
\end{align}
Consequently, the global metrics take the following functional forms
\begin{equation}\label{eq:C_grid}
\mathcal{C}(\mathcal{R},p_{\rm min}) = \frac{\mathrm{TP}(\mathcal{R},p_{\rm min})} {\mathrm{TP}(\mathcal{R},p_{\rm min}) + \mathrm{FN}(\mathcal{R},p_{\rm min})}\,,
\end{equation}
\begin{equation}\label{eq:P_grid}
\mathcal{P}(\mathcal{R},p_{\rm min}) = \frac{\mathrm{TP}(\mathcal{R},p_{\rm min})} {\mathrm{TP}(\mathcal{R},p_{\rm min}) + \mathrm{FP}(\mathcal{R},p_{\rm min})}\,,
\end{equation}
\begin{equation}\label{eq:F_grid}
\mathcal{F}(\mathcal{R},p_{\rm min}) = \frac{\mathrm{FP}(\mathcal{R},p_{\rm min})} {\mathrm{FP}(\mathcal{R},p_{\rm min}) + \mathrm{TN}(\mathcal{R},p_{\rm min})}\,.
\end{equation}
Equations \eqref{eq:C_grid}--\eqref{eq:F_grid} represent the global completeness, purity, and failure rate, respectively. Similarly to the \texttt{LinearClassification} class (Sect.\ \ref{sec:method:linear}), the \texttt{process} method of the \texttt{GridClassification} class implements an algorithm that maximizes the global completeness given a global purity threshold, denoted as $\mathcal{P}_{\rm min}$. The user must set $\mathcal{P}_{\rm min}$, along with the step size $\delta p_{\rm step}$ and the parameter $\alpha$ defining the significance of color--color bins (Appendix \ref{app:significance}). The calibration algorithm executes the following steps.
\begin{enumerate}
    \item Update the iterative probability threshold $p_{\rm min}^n=1 - n\,\delta p_{\rm step}$, where $n$ is the iteration index, starting from 0.
    \item In each $(x,y)$ space, compute $\mathcal{M}_{\rm sel}(\mathcal{R},\delta x, \delta y,p_{\rm min}^n)$ (Eq.\ \ref{eq:M_selection}). 
    \item Compute the global purity $\mathcal{P}(\mathcal{R},p_{\rm min}^n)$ (Eq.\ \ref{eq:P_grid}), assuming that all color--color bins yield a significant measurement of $p_{\rm sel}$ (Eq.\ \ref{eq:p_selection}).
    \item If $\mathcal{P}(\mathcal{R},p_{\rm min}^n)>\mathcal{P}_{\rm min}$, repeat from point 1.
\end{enumerate}
As discussed in Appendix \ref{app:significance}, the significance of $p_{\rm sel}$ measurements is assessed through a Bayesian approach. Not significant bins are masked at the end of the calibration algorithm detailed above. Then, the selection metrics (Eqs.\ \ref{eq:C_grid}--\ref{eq:F_grid}) are updated. The output of this process is a set of $N_{\rm space}$ optimal masks, where $N_{\rm space}$ is the number of color--color spaces considered for the calibration. As shown in Fig.\ \ref{fig:flowchart}, these calibrated masks can be passed directly to the \texttt{Classifier} class or saved to a \texttt{FITS} file for later use, allowing the selection to be applied to a test dataset. Analogously to the \texttt{LinearClassification} class (Sect.\ \ref{sec:method:linear}), the \texttt{GridClassification} class derives uncertainties on $\mathcal{C}$, $\mathcal{P}$, and $\mathcal{F}$ through bootstrap resampling. \\
\indent Compared to linear color cuts, the calibration algorithm presented in this section does not explicitly maximize completeness. Instead, completeness is maximized as a consequence of identifying the largest possible color--color regions where $p_{\rm sel}$ is maximal. Another conceptual difference between the methods implemented in \texttt{LinearClassification} and \texttt{GridClassification} is reported in Fig.\ \ref{fig:conceptual_diff}. As discussed in Sect.\ \ref{sec:method:linear}, linear color cuts are suited for selecting galaxies above a given redshift, as they lie at the boundaries of color--color distributions. The \texttt{GridClassification} methods, by contrast, aim to retrieve samples whose colors lie in the bulk of the color--color spaces. \\
\indent Beyond binary selection, \texttt{GridClassification} plays a key role in the probabilistic methods implemented in \texttt{FOMO-Z}: it provides the basis for the feature-selection procedure used by the \texttt{PointProbability}, \texttt{ProbabilityFunction}, and \texttt{Reconstruction} classes described in the following sections. This procedure uses the purity and completeness of calibrated selections to identify relevant colors, connecting binary selection and probabilistic inference within a common framework. This feature-selection algorithm is detailed in Appendix~\ref{app:features}.

\subsection{\texttt{\textup{PointProbability}}: Point-estimate probabilities}\label{sec:method:point_prob}
The \texttt{LinearClassification} and \texttt{GridClassification} classes calibrate binary galaxy selections, aimed at maximizing completeness given a prescribed minimum purity. This approach is particularly useful when the scientific objective is to construct a sample satisfying a given criterion, without requiring individual probability estimates. For example, weak lensing background selection can be addressed through simple, interpretable color cuts using \texttt{LinearClassification}, while \texttt{GridClassification} allows for more flexible selection boundaries in color–color spaces. For applications requiring probabilistic information for individual galaxies, \texttt{PointProbability}, described in this section, estimates the probability that an object satisfies specified selection criteria, whereas \texttt{ProbabilityFunction} (Sect.~\ref{sec:method:prob_func}) estimates the probability density function of a chosen galaxy property, such as redshift. The \texttt{probability} subpackage also includes \texttt{Reconstruction}, which estimates property distributions of selected galaxy populations (Sect.~\ref{sec:method:reconstruction}). \\
\indent The \texttt{PointProbability} method identifies reference objects whose colors are compatible with those of a test object within the adopted photometric uncertainties. For a test object indexed by $i$, we define the set of color-compatible reference objects as
\begin{equation}\label{eq:set_similar}
\Sigma_i =
\left\{
j \in \Phi :
\left|c_{i,k}-c_{j,k}\right| \leq \epsilon_{ij,k}
\quad \forall k \in \chi_i
\right\}\,.
\end{equation}
Here, $\Phi$ denotes the reference sample, while $c_{i,k}$ and $c_{j,k}$ are the values of color $k$ for the test and reference objects, respectively. The object-specific set $\chi_i \subseteq X_{\mathrm{sel}} \subseteq X$ contains the selected colors for which both a measurement and its uncertainty are available for the test object $i$. Here, $X$ denotes the full set of available colors (see Sect.\ \ref{sec:method:dataset}), and $X_{\mathrm{sel}}$ is the subset retained by the feature-selection procedure described in Appendix~\ref{app:features}. Such a feature selection identifies colors that distinguish reference objects satisfying the target criteria from those that do not. If feature selection is not performed, $X_{\mathrm{sel}}=X$. \\
\indent In words, Eq.\ \eqref{eq:set_similar} defines the set of reference objects whose colors differ from those of test object $i$ by no more than the corresponding matching tolerance in every color included in $\chi_i$. In the current implementation, the matching tolerance in Eq.\ \eqref{eq:set_similar} is expressed as
\begin{equation}\label{eq:tolerance}
\epsilon_{ij,k} =
\sqrt{\sigma_{i,k}^2 + \sigma_{j,k}^2}\,,
\end{equation}
where $\sigma_{i,k}$ and $\sigma_{j,k}$ denote the photometric uncertainties on color $k$ for the test and reference objects, respectively. This expression assumes uncorrelated photometric errors between the test and reference measurements. Furthermore, the covariance between colors sharing a band is not modeled in the current version of the code. If no test sample is supplied, or if the same \texttt{Dataset} instance is explicitly supplied as both reference and test sample, the matching tolerance is instead defined as $\epsilon_{ij,k}=\sigma_{i,k}$, without adding the reference uncertainty in quadrature. \\
\indent For a test object with a nonempty compatible reference set $\Sigma_i$, the probability of satisfying the selection criteria $\mathcal{R}$ (Eq.\ \ref{eq:R}) is estimated as
\begin{equation}\label{eq:point_probability}
p_i(\mathcal{R}) =
\frac{\sum_{j\in\Sigma_i} w_{\rm sel}^j(\mathcal{R})}
     {\sum_{j\in\Sigma_i} w_{\rm prior}^j}\,,
\end{equation}
where $w_{\rm prior}^j$ is the prior weight of the reference object $j$, and $w_{\rm sel}^j(\mathcal{R})$ is its selection weight (Eq.\ \ref{eq:w_selection}). The selection weight combines the reference prior weight with its membership in the target population, using reference PDZs for redshift criteria when available (Sect.~\ref{sec:method:sel}). In Sect.\ \ref{sec:results:quiescent}, we present a case study using this method to estimate, for each test galaxy, its probability of being star-forming. \\
\indent We emphasize that photometric uncertainties explicitly set the matching tolerance used to define $\Sigma_i$, whereas intrinsic scatter in the color--property relation enters through the reference sample itself. Consequently, even objects with closely matching colors may differ in whether they satisfy $\mathcal{R}$. The probability in Eq.~\eqref{eq:point_probability} is therefore determined by the weighted target memberships within a neighborhood whose extent depends on the photometric uncertainties, while the diversity of reference properties captures the scatter in the color--property relations.

\subsection{\texttt{\textup{ProbabilityFunction}}: Probability distributions}\label{sec:method:prob_func}
\begin{figure}[t]
\centering\includegraphics[width = \hsize-1cm, height = 4cm] {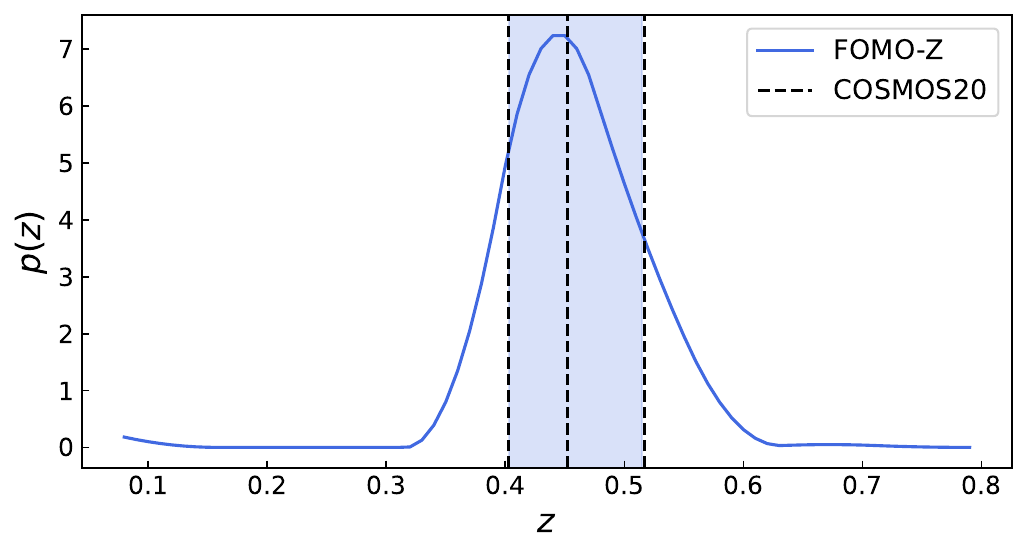}
\caption{Example redshift PDF reconstructed by \texttt{ProbabilityFunction} (Sect.\ \ref{sec:method:prob_func}) for a COSMOS20 galaxy, using COSMOS20 as the reference sample. The blue curve shows the reconstructed PDF, and the blue shaded region marks its 68\% confidence interval. The black dashed vertical lines mark the 16th, 50th, and 84th percentiles of the COSMOS20 redshift PDF, from left to right. This example illustrates the reconstruction rather than validation on an independent test sample.}
\label{fig:prob_func_1}
\end{figure}
\begin{figure}[t]
\centering\includegraphics[width = \hsize-2cm, height = 7cm] {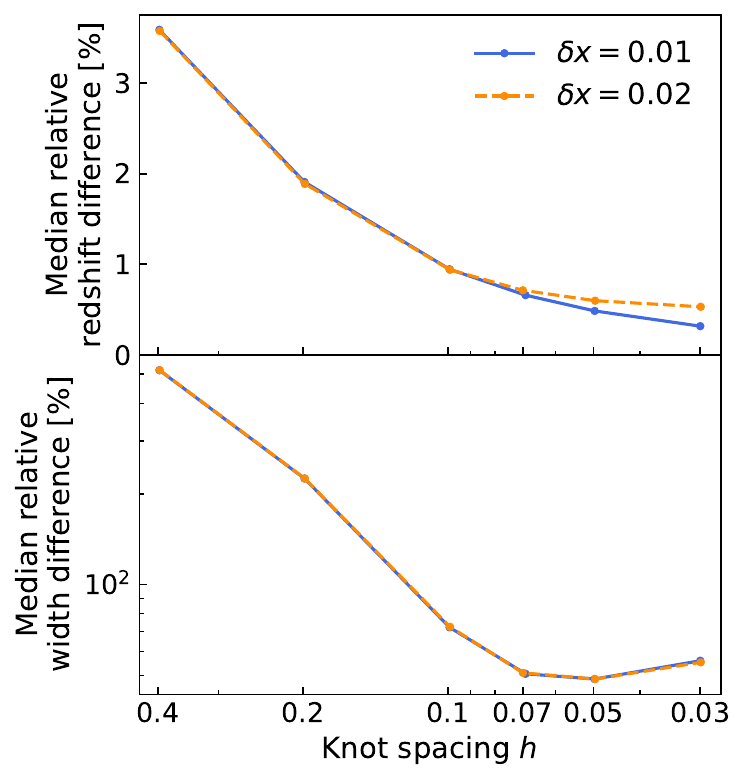}
\caption{Dependence of the reconstructed redshift PDFs, derived using the \texttt{ProbabilityFunction} methods (Sect.\ \ref{sec:method:prob_func}), on the spline knot spacing $h$, for CDF threshold spacings $\delta x=0.01$ (blue solid lines) and $0.02$ (orange dashed lines). Top: median absolute relative difference between the reconstructed median redshift and the COSMOS20 catalog estimate, $\mathrm{median}\!\left( |z_{50}^{\scriptscriptstyle\rm FOMO-Z}-z^{\scriptscriptstyle\rm COSMOS20}|/ z^{\scriptscriptstyle\rm COSMOS20}\right)$. Bottom: median absolute relative difference in the widths of the 68\% intervals, $\mathrm{median}\!\left( |W^{\scriptscriptstyle\rm FOMO-Z}-W^{\scriptscriptstyle\rm COSMOS20}|/ W^{\scriptscriptstyle\rm COSMOS20}\right)$, where $W^{\scriptscriptstyle\rm FOMO-Z}=z_{84}^{\scriptscriptstyle\rm FOMO-Z}-z_{16}^{\scriptscriptstyle\rm FOMO-Z}$ and $W^{\scriptscriptstyle\rm COSMOS20}$ is the difference between the catalog's upper and lower 68\% bounds. Both statistics are expressed as percentages. The horizontal axis shows the knot spacing on a logarithmic scale, decreasing toward the right. COSMOS20 serves as both the reference and test sample, so this comparison assesses consistency with the catalog information.}
\label{fig:prob_func_2}
\end{figure}

The \texttt{PointProbability} class described in Sect.\ \ref{sec:method:point_prob} estimates the probability that a test object satisfies user-defined selection criteria. For example, it can estimate the probability that object $i$ is a quasar or that its stellar mass exceeds a specified threshold. For a continuous property, such as redshift, evaluating these probabilities across a range of thresholds provides the basis for estimating its probability density function (PDF) for each test object. This procedure is implemented in the \texttt{ProbabilityFunction} class.\\
\indent For a test galaxy $i$, we reconstruct the PDF, $f_i(x)$, of a continuous property $x$ by first estimating its cumulative distribution function (CDF),
\begin{equation}\label{eq:cdf_definition}
F_i(t) = \int_{-\infty}^{\,t} f_i(x)\,{\rm d}x\,,
\end{equation}
which gives the probability that the property lies below a threshold $t$. This probability can be estimated directly using the \texttt{PointProbability} class described in Sect.\ \ref{sec:method:point_prob}. Specifically, we define the threshold-dependent selection criterion
\begin{equation}\label{eq:cdf_rules}
\mathcal{R}_t : x<t\,.
\end{equation}
The corresponding sampled CDF estimate is \begin{equation}\label{eq:cdf_point_probability} \widetilde{F}_i(t) = p_i(\mathcal{R}_t)\,, \end{equation} where the right-hand side is computed using Eq.~\eqref{eq:point_probability}. Repeating this calculation over a set of thresholds provides sampled estimates of the CDF from which the PDF can be reconstructed. \\
\indent Feature selection is performed separately at each threshold $t$ using the procedure described in Appendix~\ref{app:features}. The retained colors may therefore vary with $t$, reflecting changes in their ability to distinguish objects on either side of the threshold. For example, different colors may be informative at different redshifts as spectral features shift across the observed bands. Consequently, the set of color-compatible reference objects (see Eq.\ \ref{eq:set_similar}) may also vary with the threshold, and we denote it here by $\Sigma_i(t)$. Because different thresholds can use different reference subsets, the sampled estimates $\widetilde{F}_i(t)$ are not guaranteed to be non-decreasing. \\
\indent To obtain a smooth, non-decreasing representation of the CDF, we fit the sampled estimates with a cubic B-spline over the sampled threshold interval $[t_{\min},t_{\max}]$. The user specifies a knot spacing $h$, expressed in the units of the reconstructed property. We divide the interval into
\begin{equation}
N_{\rm int} =
{\rm int}\left( \frac{t_{\max}-t_{\min}}{h} \right)
\end{equation}
equal subintervals. This gives $N_{\rm knots}=N_{\rm int}-1$ internal knots, excluding the two boundaries of the sampled interval.\footnote{The full knot vector includes the internal knots and the two boundary values, $t_{\min}$ and $t_{\max}$, both repeated four times. For a cubic spline, this defines the basis over the full interval without requiring external knots and gives $\widehat{F}_i(t_{\min})=a_{i,0}$ and $\widehat{F}_i(t_{\max})=a_{i,M-1}$. The probability assigned to this interval is therefore simply $a_{i,M-1}-a_{i,0}$. These boundary probabilities remain free parameters of the fit. They are not fixed to zero and one, since the sampled interval may not contain the full probability distribution.} The fitted CDF takes the form
\begin{equation}\label{eq:cdf_spline}
\widehat{F}_i(t)=\sum_{m=0}^{M-1}a_{i,m}B_{m,3}(t)\,,
\end{equation}
where the cubic B-spline basis functions $B_{m,3}$ are fixed by the knot positions, and $M=N_{\rm knots}+4=N_{\rm int}+3$ is the number of basis functions. The coefficients $a_{i,m}$ are fitted separately for each test object $i$ through constrained least-squares minimization. We restrict the coefficients to $[0,1]$ to ensure that the fitted CDF remains within the valid probability range. In fact, the B-spline basis functions are non-negative and sum to unity over the fitted interval. We additionally impose the sufficient condition for monotonicity
\begin{equation}\label{eq:cdf_monotonicity}
a_{i,m+1}\geq a_{i,m}
\qquad m=0,\ldots,M-2\,.
\end{equation}
The derivative of the fitted cubic spline is a linear combination of non-negative quadratic B-spline basis functions. Each coefficient in this combination is a positive multiple of the difference $a_{i,m+1}-a_{i,m}$. The constraint in Eq.\ \eqref{eq:cdf_monotonicity} therefore makes every term in the derivative non-negative, ensuring that ${\rm d}\widehat{F}_i/{\rm d}t\geq0$ throughout the sampled interval. The derivative therefore defines a non-negative density over the
sampled interval. Provided that
$\widehat{F}_i(t_{\max})>\widehat{F}_i(t_{\min})$,
the normalized PDF is
\begin{equation}\label{eq:pdf_spline}
\widehat{f}_i(x)=
\frac{{\rm d}\widehat{F}_i(x)/{\rm d}x}
{\widehat{F}_i(t_{\max})-\widehat{F}_i(t_{\min})}\,,
\qquad x\in[t_{\min},t_{\max}]\,.
\end{equation}
The resulting PDF integrates to unity within the sampled interval. In the code implementation, the analytic spline derivative is evaluated at the sampled thresholds, and the resulting density samples are normalized using trapezoidal quadrature. This provides a numerical approximation to Eq.~\eqref{eq:pdf_spline}, whose accuracy depends on the threshold spacing. \\
\indent Figure~\ref{fig:prob_func_1} shows the application of \texttt{ProbabilityFunction} to a COSMOS20 galaxy, showing its reconstructed redshift PDF. COSMOS20 also serves as the reference sample for this illustrative example, using its redshift point estimates without their associated uncertainties. We remark that these reconstructions depend on two parameters: the spacing $\delta x$ between consecutive thresholds at which the CDF is estimated, and the maximum knot spacing $h$ of the fitted spline. Smaller values of $\delta x$ provide finer sampling of the CDF, while smaller values of $h$ allow for more sensitivity to small variations in the PDF. The choice of these parameters therefore requires balancing flexibility against sensitivity to fluctuations in the estimated cumulative probabilities.\\
\indent We use photometric redshift estimation as a case study to
investigate this dependence. Using COSMOS20 (Sect.~\ref{sec:data})
photometry in nine bands ($g,r,i,z,B,Y,J,H,K_{\rm s}$), we reconstruct
individual redshift PDFs for $\delta x=0.01$ and $0.02$ over
$0.02\leq z<4.8$, and knot spacings
$h=0.4,\,0.2,\,0.1,\,0.07,\,0.05,$ and $0.03$.
For each $\delta x$, we refit the same cumulative probability estimates
with different values of $h$.
Figure~\ref{fig:prob_func_2} compares the resulting medians and central
68\% probability intervals with the COSMOS20 catalog redshifts and their
reported 68\% intervals. Since the same catalog serves as both the
reference and test sample, this exercise assesses consistency with the
reference information rather than predictive performance on an
independent sample. The two sampling intervals, $\delta x=0.01$ and 0.02, yield similar results. On the other hand, the choice of $h$ has a more pronounced effect. Smaller knot
spacings improve agreement with catalog estimates in the explored range. In general, the reconstructed medians and 68\% intervals agree
well with the COSMOS20 estimates in this consistency test. The reconstructions obtained in this example use reference redshift point estimates
without propagating their associated uncertainties. Incorporating the
full reference PDZs is possible, as detailed above, and would improve the quality of the reconstructed photo-$z$s. Alternatively, a spectroscopic reference sample would provide more precise redshift information, provided that it adequately represents the target population. Lastly, the median differences shown in Fig.~\ref{fig:prob_func_2} summarize the overall agreement but do not show how it varies with redshift. A more complete assessment should therefore evaluate these differences as a function of redshift for the reference and test samples used.

\subsection{\texttt{\textup{Reconstruction}}: Reconstructed sample distributions}\label{sec:method:reconstruction}
The \texttt{PointProbability} class (Sect.\ \ref{sec:method:point_prob}) matches reference and test objects based on their photometric properties (Eq.\ \ref{eq:set_similar}) and assigns membership probabilities to the test objects (Eq.\ \ref{eq:point_probability}). Building on the same matching framework, the \texttt{Reconstruction} class infers population-level distributions of properties in a selected test sample using measurements available in the reference sample. \\
\indent The method takes a reference and a test sample, together with an observed property $s$ available in both samples and an interval defining the target population:
\begin{equation}\label{eq:reconstruction_selection}
\mathcal{R}_{\rm obs} :
s_{\rm min} < s \leq s_{\rm max}\,.
\end{equation}
The property $s$ used for this selection should not coincide with the property $x$ whose distribution is to be reconstructed. For example, a tomographic sample may be selected using photometric redshift estimates, while its underlying redshift distribution is inferred using spectroscopic measurements in the reference sample. The selection in Eq.~\eqref{eq:reconstruction_selection} is applied to observed values in both the reference and test samples.\\
\indent The selection rule $\mathcal{R}_{\rm obs}$ also defines the identification of informative colors. The feature-selection procedure described in Appendix~\ref{app:features} uses the full reference sample to identify colors that best distinguish objects satisfying $\mathcal{R}_{\rm obs}$ from those outside it. The reference sample must therefore be supplied before applying this interval selection. Matching is subsequently performed between the selected test subset, $\Psi_{\mathcal{R}_{\rm obs}}$, and the selected reference subset, $\Phi_{\mathcal{R}_{\rm obs}}$. For each selected test object $i$, the set $\Sigma_i$ of compatible reference objects is defined by Eq.~\eqref{eq:set_similar}, with $\Phi$ replaced by $\Phi_{\mathcal{R}_{\rm obs}}$. A selected test object contributes to the reconstruction only if
it has at least the required number of usable selected colors
(two by default) and at least one compatible reference object with
positive prior weight. \\
\indent Each selected test object $i$ contributes its weight $w_{\rm test}^i$, which is distributed among its compatible reference objects. The total reconstruction weight transferred to reference object $j$ is therefore
\begin{equation}\label{eq:reconstruction_weights}
w_{\rm rec}^j =
\sum_{\substack{
i\in\Psi_{\mathcal{R}_{\rm obs}}\\
\text{such that }j\in\Sigma_i
}}
w_{\rm test}^i
\frac{w_{\rm prior}^j}
{\sum_{k\in\Sigma_i}w_{\rm prior}^k}\,,
\end{equation}
where $\Psi_{\mathcal{R}_{\rm obs}}$ denotes the subset of test objects that satisfy $\mathcal{R}_{\rm obs}$. The weights $w_{\rm test}^i$ specify each test object's contribution to the target population, and default to unity. For example, shape-measurement weights can be used when reconstructing the redshift distribution relevant to a cosmic shear analysis. The reference prior weights $w_{\rm prior}^k$, instead, determine the relative contributions of compatible reference analogs. For a fixed reference object $j\in\Phi_{\mathcal{R}_{\rm obs}}$, the sum in Eq.\ \eqref{eq:reconstruction_weights} includes only selected test objects $i$ whose compatible reference set $\Sigma_i$ contains $j$. For each such test object, the denominator sums the prior weights of all reference objects $k\in\Sigma_i$. This ensures that each contributing test object transfers a total weight equal to $w_{\rm test}^i$, independently of the number of compatible reference objects. \\
\indent For illustration, if all reference prior weights and test weights are unity, Eq.~\eqref{eq:reconstruction_weights} reduces to
\begin{equation}
w_{\rm rec}^j =
\sum_{\substack{
i\in\Psi_{\mathcal{R}_{\rm obs}}\\
\text{such that }j\in\Sigma_i
}}
\frac{1}{|\Sigma_i|}\,,
\end{equation}
where $|\Sigma_i|$ is the number of compatible reference objects for test object $i$. For example, consider two test objects with unit weights and unit reference prior weights: the first has 10 compatible reference objects, while the second has 100. The first distributes a weight of $1/10$ to each of its analogs, and the second distributes $1/100$. Each test object therefore contributes a total weight of unity: $10\times(1/10)=100\times(1/100)=1$. A reference object compatible with both receives $w_{\rm rec}^j=1/10+1/100=0.11$. If each match instead contributed unit weight, the two test objects would contribute total weights of 10 and 100, respectively, despite having identical test weights $w_{\rm test}$. This is why we normalize each test object's contribution by the total prior weight of its compatible reference objects: its contribution to the reconstructed distribution should depend on its assigned test weight, rather than on how many reference analogs are available. \\
\indent The distribution of property $x$ in the selected test sample is estimated by binning the reference measurements with their transferred weights $w_{\rm rec}^j$:
\begin{equation}\label{eq:reconstruction_counts}
N_b =
\sum_{j\in\Phi_{\mathcal{R}_{\rm obs}}}
w_{\rm rec}^j\,\delta^j(x\in B_b)\,,
\end{equation}
Here, $B_b=[e_b,e_{b+1})$ is the bin with index $b$, while $\delta^j(x\in B_b)$ equals unity if the reference value $x_j$ lies within bin $B_b$, and zero otherwise. The corresponding normalized probability density is
\begin{equation}\label{eq:reconstruction_density}
n_b =
\frac{N_b}{(e_{b+1}-e_b)\sum_c N_c}\,.
\end{equation}
The density is normalized over the provided bin range. \\
\indent When reconstructing redshift distributions using reference PDZs, the binary bin membership in Eq.~\eqref{eq:reconstruction_counts} is replaced by the probability mass within each bin:
\begin{equation}\label{eq:reconstruction_pdz}
N_b =
\sum_{j\in\Phi_{\mathcal{R}_{\rm obs}}}
w_{\rm rec}^j
\int_{e_b}^{e_{b+1}} p^j(z)\,\mathrm{d}z\,.
\end{equation}
This follows the same treatment of reference redshift uncertainty introduced in Eq.~\eqref{eq:R_z}. Reference PDZs enter this final distribution estimate, whereas the initial sample selection $\mathcal{R}_{\rm obs}$ remains based on observed point values.

\section{Applications}\label{sec:results}
\begin{figure}[t]
\centering\includegraphics[width = \hsize, height = 6cm] {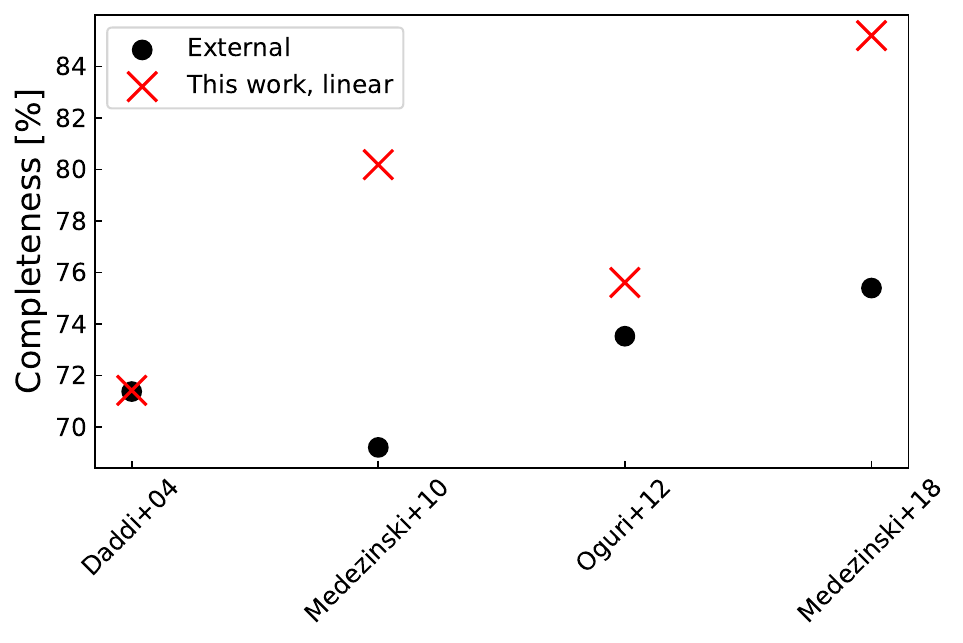}
\caption{Comparison of selection completeness between literature background selections (black dots) and \texttt{FOMO-Z} (red crosses) for four literature references: \citet{Daddi04}, \citet{Medezinski10}, \citet{Oguri12}, and \citet{Medezinski18} (left to right). The \texttt{FOMO-Z} results are derived using the linear color cut method detailed in Sect.\ \ref{sec:method:linear}. To enable a fair comparison, we applied each literature selection to the COSMOS20 galaxy sample, computed the corresponding purity, and used it as the threshold in \texttt{FOMO-Z}, while also adopting the same photometric bands and limiting redshifts as the original study.}
\label{fig:external_comparison}
\end{figure}
\begin{figure*}[ht]
\centering\includegraphics[width = \hsize, height = 6cm] {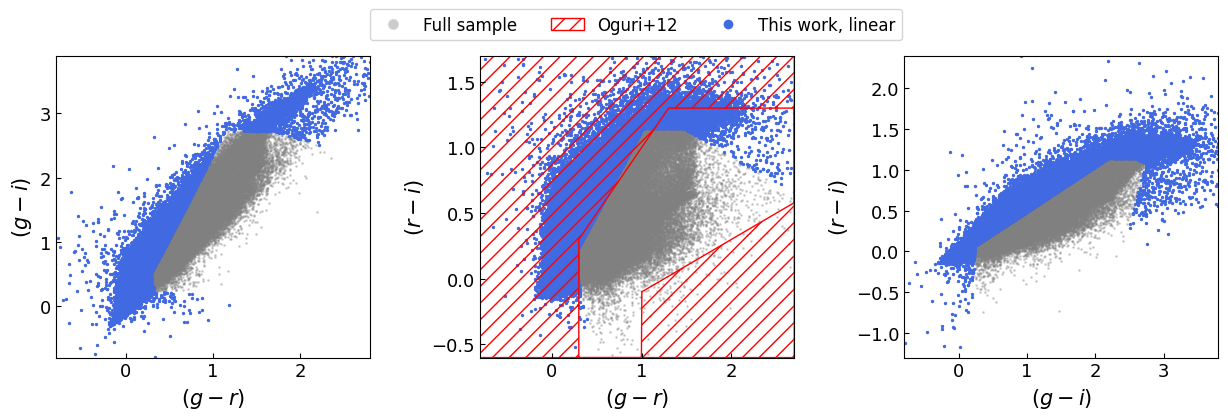}
\caption{Comparison between the calibration of linear color conditions described in Sect.~\ref{sec:method:linear} and the selection proposed by \citet{Oguri12}. Gray points show the full COSMOS20 galaxy sample, while blue points indicate the galaxies selected by \texttt{FOMO-Z}. The dashed red regions in the middle panel mark the color selection criteria of \citet{Oguri12}. To enable a fair comparison, the \texttt{FOMO-Z} calibration was performed using the same photometric bands and targeting the same purity level obtained by applying the \citet{Oguri12} selection to the COSMOS20 sample.}
\label{fig:oguri12}
\end{figure*}
In the following, we present example applications of the methods implemented in the \texttt{FOMO-Z} code, and compare our results with similar classifications and property reconstructions from the literature. In Sect.~\ref{sec:results:background}, we construct background galaxy selections; Sect.\ \ref{sec:results:maximize} presents a method to maximize the weak lensing S/N using \texttt{FOMO-Z}; in Sect.~\ref{sec:results:quiescent}, we derive selections of quiescent and star-forming galaxies; and in Sect.~\ref{sec:results:soms}, we reconstruct galaxy redshift distributions in tomographic bins and compare them with results from self-organizing maps. A detailed scientific interpretation of the results presented in the following is beyond the scope of this paper.

\subsection{High-redshift galaxies}\label{sec:results:background}
In this section, we compare high-redshift galaxy selections from \texttt{FOMO-Z} with analogous ones from the literature. To define such background selections, we formally express $\mathcal{R}_z$ in Eq.\ \eqref{eq:zl} as follows:
\begin{equation}
\mathcal{R}_z : z_{\rm g} \in [z_{\rm l}, \infty]\,,
\end{equation}
for arbitrary choices of $z_{\rm l}$. Background source color selections have traditionally been defined via empirical, hand-tuned cuts in color–color spaces. \citet{Oguri12}, for instance, construct their background sample from the COSMOS catalog of \citet{Ilbert09}, by visually identifying the regions in the $g-r$ versus $r-i$ plane that are predominantly populated by galaxies at $z\gtrsim0.7$. The boundaries are set by eye, without an explicit statistical criterion. This approach works, but it is time-consuming, survey-dependent, and difficult to reproduce or port to other datasets. \texttt{FOMO-Z} fully automates this step, yielding reproducible selections that can be recalibrated automatically for any survey, filter set, or depth with no manual tuning required. \\
\indent Similar approaches, validated with theoretical galaxy evolutionary tracks, were adopted by \citet{Medezinski10}, who calibrated a background selection based on $gri$ photometry for clusters spanning $0.26<z<0.45$, by \citet{Medezinski18}, who derived a background selection for clusters in the redshift range $0.1<z<1.1$ using $griz$ bands, and by \citet{Daddi04}, who selected galaxies at $z>1.4$ using $BzK$ photometry. \\
\indent To compare the \texttt{FOMO-Z} performance with these literature selections, we used the linear color cut method described in Sect.\ \ref{sec:method:linear}, which is best suited for defining background galaxy samples. We employed the COSMOS20 catalog introduced in Sect.\ \ref{sec:data}, considering $griz$ photometry from HSC, the $U$ band from the Subaru Suprime-Cam \citep{Capak07,Taniguchi15}, and the $K_s$ band from UltraVISTA \citep{McCracken12,Moneti19}. To make a fair comparison, we applied the literature selections to this sample, derived the corresponding purity values at the relevant lower redshift limits $z_{\rm l}$, and used these as purity thresholds in \texttt{FOMO-Z}. Applying the \citet{Oguri12} selection at $z_{\rm l}=0.65$ yields a purity of $\mathcal{P}=0.96$. The \citet{Medezinski10} selection at $z_{\rm l}=0.26$ gives $\mathcal{P}=0.97$, the \citet{Medezinski18} selection at $z_{\rm l}=0.8$ gives $\mathcal{P}=0.90$, and the \citet{Daddi04} selection at $z_{\rm l}=1.4$ yields $\mathcal{P}=0.91$. In addition, in \texttt{FOMO-Z} we adopted the same sets of filters used to derive the literature selections. \\
\indent Figure \ref{fig:external_comparison} displays the comparison with \texttt{FOMO-Z} results based on the same $z_{\rm l}$ values and sets of photometric bands. Under these conditions, \texttt{FOMO-Z} yields equal or higher completeness than the literature selections. To illustrate the performance of our methods in color--color spaces, Fig.\ \ref{fig:oguri12} compares the color distributions from the \citet{Oguri12} selection with those from \texttt{FOMO-Z}, assuming the same calibration setup. The middle panel in the figure shows that \texttt{FOMO-Z} reproduces well the color cuts derived by \citet{Oguri12} in the $(g-r)$--$(r-i)$ plane, except in the bottom-right region, where galaxies are excluded due to poor statistics. Figure~\ref{fig:oguri12} highlights an additional advantage of \texttt{FOMO-Z}: it does not require prior knowledge of the most informative color spaces. As described in Sect.~\ref{sec:method:dataset}, the analysis explores all possible color combinations derived from the available photometric bands. Nevertheless, the user may optionally restrict the analysis to a subset of color spaces (Sect.~\ref{sec:method:dataset}). We verified that limiting the calibration to the $(g-r)$--$(r-i)$ color space does not significantly affect the resulting completeness or the optimized color cut parameters. \\
\indent All in all, \texttt{FOMO-Z} proves to be more flexible and to achieve better performance than existing methods for background selections. To further illustrate the simplicity of its application, Fig.~\ref{fig:code_oguri} in Appendix~\ref{app:usage} presents a short code snippet that reproduces the results shown in Fig.~\ref{fig:oguri12}.

\subsection{Maximizing the weak lensing S/N of halo profiles}\label{sec:results:maximize}
\begin{figure}[t]
\centering\includegraphics[width = \hsize, height = 8.2cm] {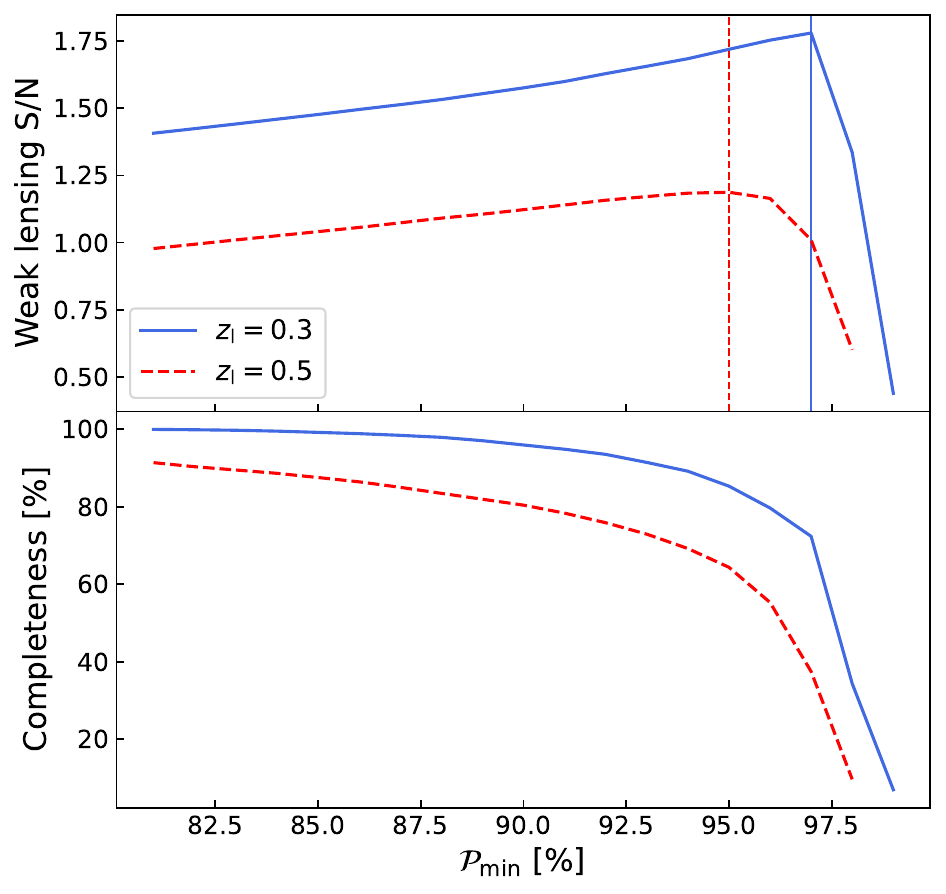}
\caption{S/N of a halo weak lensing profile (top) and completeness of the background selection (bottom) as a function of the purity threshold adopted in the color selection calibration, $\mathcal{P}_{\rm min}$. A halo with mass $M_{200\rm m}=10^{14}$ M$_\odot$ is considered, at redshifts $z_{\rm l}=0.3$ (solid blue lines) and $z_{\rm l}=0.5$ (dashed red lines). The vertical lines in the top panel are in correspondence of the maximum S/N.}
\label{fig:wl_maximization}
\end{figure}
The flexibility of the \texttt{LinearClassification} class, together with its excellent performance (see Sect.~\ref{sec:results:background} and Appendix~\ref{app:performance}), enables an efficient maximization of the S/N of weak lensing profiles. This is achieved by calibrating the color selection for a given lens redshift $z_{\rm l}$ while varying the assumed purity threshold $\mathcal{P}_{\rm min}$. \\
\indent To this end, we model the weak lensing signal of a halo as the sum of a 1-halo term, described by the \citet*{BMO} profile, and a cosmological 2-halo term \citep{Oguri11}. Specifically, we consider the reduced tangential shear, $g_{\rm t}$, expressed as follows:
\begin{equation}\label{eq:g_cen}
g_{\rm t}(R,M,z_{\rm l}) = \frac{\mathcal{P}\,\Delta\Sigma(R,M,z_{\rm l})\,\langle\Sigma_{\rm crit}^{-1}(z_{\rm l})\rangle}{1 - \Sigma(R,M,z_{\rm l})\,\langle\Sigma_{\rm crit}^{-1}(z_{\rm l})\rangle^{-1}\,\langle\Sigma_{\rm crit}^{-2}(z_{\rm l})\rangle}\,,
\end{equation}
where $R$ is the physical halo-centric radius, $M$ is the mass, and $z_{\rm l}$ is the redshift. In Eq.\ \eqref{eq:g_cen}, $\mathcal{P}$ denotes the purity of the \texttt{FOMO-Z} background selection, accounting for the dilution of the weak lensing signal due to foreground contaminants \citep{Dietrich19}. Furthermore, in Eq.\ \eqref{eq:g_cen}, $\Delta\Sigma$ is the excess surface density, $\Sigma$ is the surface density, while the effective critical surface density, $\langle\Sigma_{\rm crit}^{-\eta}\rangle$, has the following expression
\begin{equation}\label{eq:sigma_crit}
\langle\Sigma_{\rm crit}^{-\eta}(z_{\rm l})\rangle = \frac{\int_{z_{\rm g}>z_{\rm l}}{\rm d}z_{\rm g}\, \Sigma_{\rm crit}^{-\eta}(z_{\rm g},z_{\rm l})\,n(z_{\rm g}\,|\,z_{\rm l})}{\int_{z_{\rm g}>z_{\rm l}}{\rm d}z_{\rm g}\, n(z_{\rm g}\,|\,z_{\rm l})}\,.
\end{equation}
Here, $\Sigma_{\rm crit}$ is the critical surface density, $z_{\rm g}$ is the background galaxy redshift, while $n(z_{\rm g}\,|\,z_{\rm l})$ is the background redshift distribution. An approximate estimate of this distribution can be obtained by applying the binary selection mask returned by \texttt{LinearClassification} to the reference sample and constructing its weighted redshift histogram. Alternatively, the \texttt{Reconstruction} class (Sect.\ \ref{sec:method:reconstruction}) estimates this distribution by transferring redshift information from the reference sample to the selected test sample through photometric matching. \\
\indent In the following, we use the KiDZ sample (Sect.\ \ref{sec:data}) as the reference. We assume that the observational error on the measured $g_{\rm t}$ profiles, $\sigma_{g_{\rm t}}$, is uniquely defined by the intrinsic shape noise, amounting to $\sigma_\epsilon=0.25$ in KiDS \citep{Li23,Wright25_cosmo}, and by the number of galaxies included in the measurement, $N_{\rm g}$:
\begin{align}\label{eq:sigma_g}
\sigma_{g_{\rm t}}(R_{\rm min},R_{\rm max},z_{\rm l}) &= 
\sqrt{\frac{\sigma_\epsilon^2}{N_{\rm g}(R_{\rm min},R_{\rm max}, z_{\rm l})}}\nonumber\\
& =\sqrt{\frac{\sigma_\epsilon^2}{\langle n_{\rm g}(z_{\rm l})\rangle\,A(R_{\rm min}, R_{\rm max}, z_{\rm l})\,\left(\frac{180 \times 60}{\pi}\right)^2}}\,.
\end{align}
Here, $R_{\rm min}$ and $R_{\rm max}$ are the minimum and maximum halo-centric distances, respectively, while $A$ is the annulus area, defined as
\begin{equation}
A(R_{\rm min}, R_{\rm max}, z_{\rm l}) = \pi\,\frac{R_{\rm max}^2-R_{\rm min}^2}{D_{\rm A}^2(z_{\rm l})}\,,
\end{equation}
where $D_{\rm A}(z_{\rm l})$ is the angular diameter distance computed at $z_{\rm l}$. In Eq.\ \eqref{eq:sigma_g}, $\langle n_{\rm g}(z_{\rm l})\rangle$ is the effective number of background galaxies per square arcminute, expressed as follows \citep{Heymans12}:
\begin{equation}\label{eq:heymans12}
\langle n_{\rm g}(z_{\rm l})\rangle = \frac{1}{\Omega}\,\frac{\left(\sum_i w_i\right)^2}{\sum_i w_i^2}\,.
\end{equation}
Here, $\Omega$ is the survey area, $i$ runs over the background galaxies selected with \texttt{FOMO-Z}, and $w_i$ is the weight of the $i$th galaxy, assumed to be the product of the volume prior weight \citep{Wright25} and the shape weight reported in the KiDZ catalog. The shape weight is measured using the \texttt{Lensfit} algorithm \citep{Miller07,Fenech17}. \\
\indent We express the weak lensing S/N as follows \citep{Umetsu20b,Umetsu20}:
\small
\begin{equation}\label{eq:SNR}
    ({\rm S/N})_{\rm\scriptscriptstyle WL}(M,z_{\rm l}) = 
    \frac{\sum_j g_{\rm t}(R_j,M,z_{\rm l}) \, \sigma^{-2}_{g_{\rm t}}(R_j,M,z_{\rm l})}
    {\sqrt{\sum_j \sigma^{-2}_{g_{\rm t}}(R_j,M,z_{\rm l})}} \,.
\end{equation} 
\normalsize
Here, $j$ runs over the halo-centric radial bins, $R_j$ is the central value of the $j$th radial bin, the signal $g_{\rm t}$ is given by Eq.\ \eqref{eq:g_cen}, and the noise $\sigma_{g_{\rm t}}$ is obtained through Eq.\ \eqref{eq:sigma_g}. We remark that Eq.\ \eqref{eq:SNR} is an approximation of the S/N, as it neglects the cross-correlation between radial bins.\\
\indent To estimate $({\rm S/N})_{\rm\scriptscriptstyle WL}$, we considered six logarithmically spaced radial bins in the range $R\in[0.4,3.5]$ $h^{-1}$Mpc and a halo mass of $M_{200\mathrm{m}}=10^{14}$ M$_\odot$. Here, $M_{200\mathrm{m}}$ is the mass enclosed within $R_{200\mathrm{m}}$, defined as the radius within which the mean density is 200 times the mean matter density of the Universe at the halo redshift. For the \texttt{FOMO-Z} calibrations of background galaxy selections, we considered the full set of photometric bands in the KiDZ sample. Based on the \texttt{LinearClassification} class, we derived the background sample purity entering Eq.~\eqref{eq:g_cen}, reconstructed the background redshift distributions required for Eq.~\eqref{eq:sigma_crit}, and computed the effective galaxy density defined in Eq.~\eqref{eq:heymans12}. \\
\indent Figure~\ref{fig:wl_maximization} shows $({\rm S/N})_{\rm\scriptscriptstyle WL}$ obtained by varying the purity threshold in \texttt{LinearClassification}. As expected, $({\rm S/N})_{\rm\scriptscriptstyle WL}$ reaches a maximum. Indeed, low-purity selections suffer from increased contamination, which dilutes the weak lensing signal (see Eq.\ \ref{eq:g_cen}), while overly strict purity requirements reduce the number of selected background galaxies and therefore the effective galaxy density. An optimal balance between purity and completeness is therefore essential to maximize the weak lensing S/N and obtain tighter constraints on halo masses and other key physical properties.

\subsection{Quiescent and star-forming galaxies}\label{sec:results:quiescent}
\begin{figure}[t]
\centering\includegraphics[width = \hsize, height = 10.5cm] {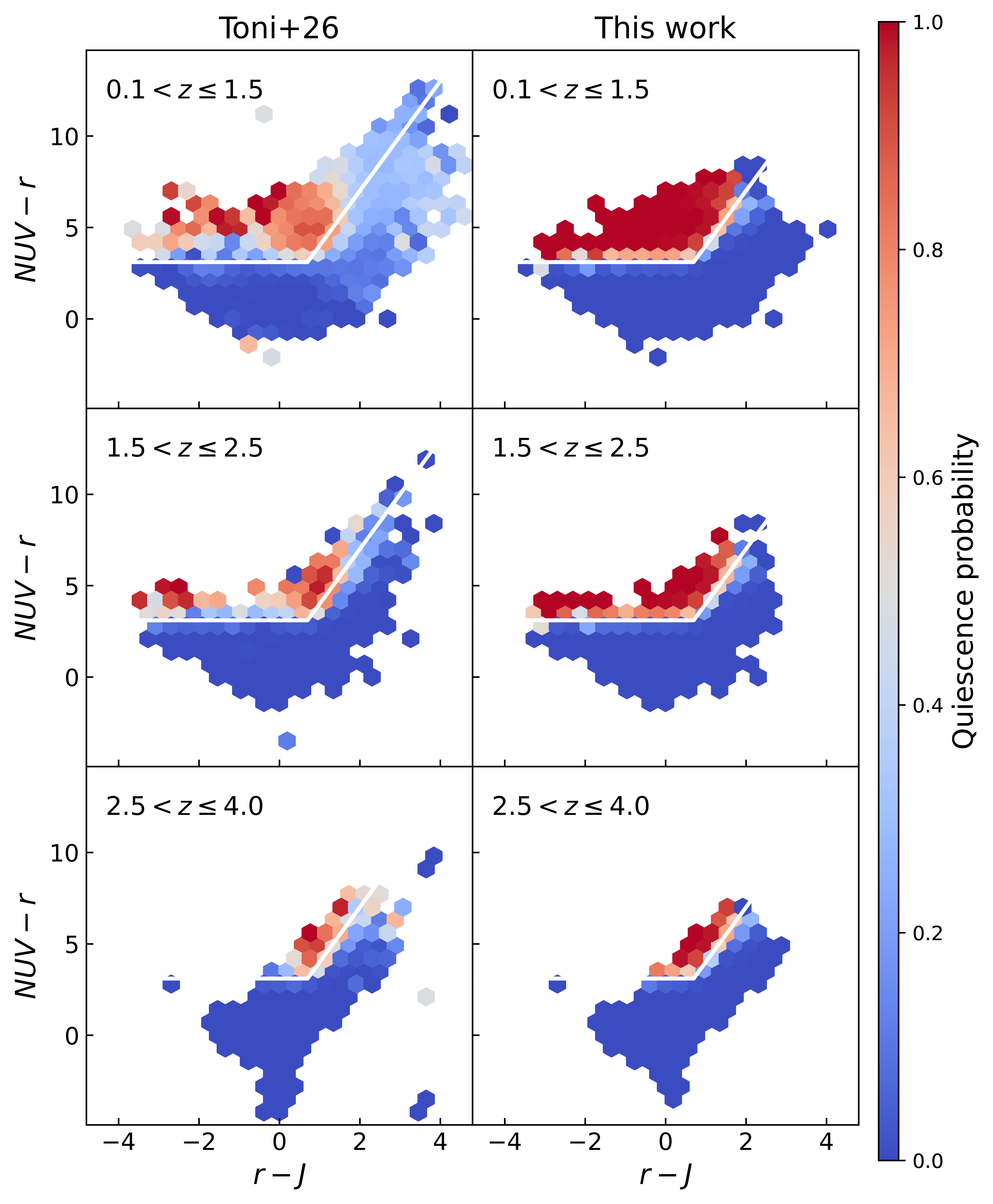}
\caption{Quiescence probabilities estimated by \citet{Toni25} using the \texttt{XGB} algorithm \citep{Chen16} (left panels) and by \texttt{FOMO-Z} using the \texttt{PointProbability} class (right panels; Sect.~\ref{sec:results:quiescent}). Each hexagonal bin is color-coded by the mean quiescence probability of the galaxies it contains. The white lines indicate the color--color cuts of \citet{Ilbert13}.}
\label{fig:Toni26_comparison}
\end{figure}
\begin{figure*}[ht]
\centering\includegraphics[width = \hsize-2cm, height = 8.1cm] {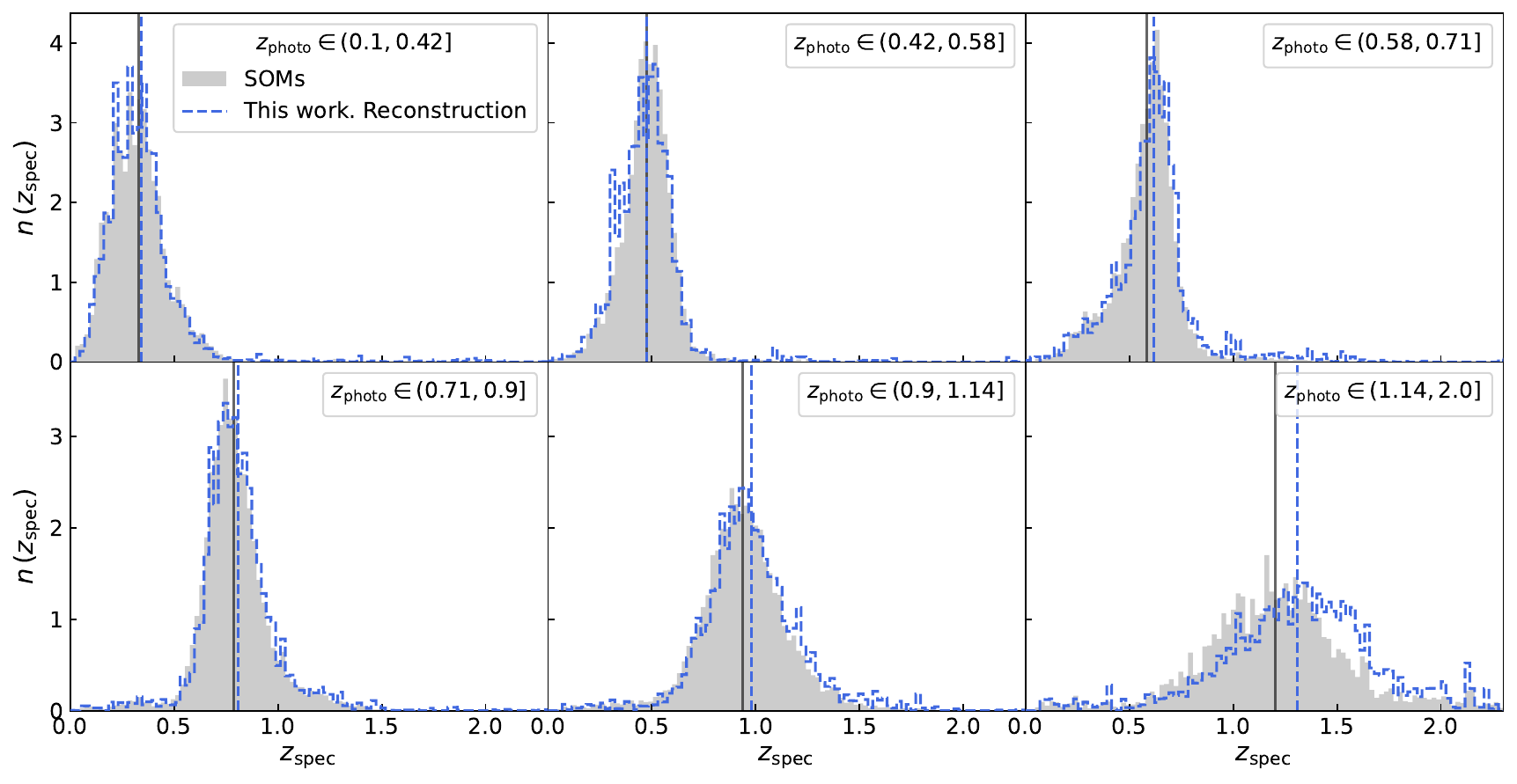}
\caption{Comparison of the normalized redshift distributions reconstructed using the SOM pipeline of \citet{Wright25} and the \texttt{Reconstruction} class of \texttt{FOMO-Z} (Sect.~\ref{sec:method:reconstruction}). Each panel corresponds to a tomographic bin defined in photometric redshift, with boundaries indicated in the legend title. Gray histograms show the SOM reconstructions, while dashed blue curves show the \texttt{FOMO-Z} results. Solid black and dashed blue vertical lines mark the corresponding weighted mean redshifts. Neither reconstruction includes simulation-based bias corrections.}
\label{fig:som}
\end{figure*}
Separating quiescent from star-forming galaxies is a key step in characterizing galaxy populations and their environments, and a variety of photometric methods have been developed for this purpose, ranging from simple color--color cuts, such as the widely used rest-frame $UVJ$ selection \citep[e.g.,][]{Williams09,Whitaker11}, to more recent machine-learning classifiers trained on multi-band photometry. \citet{Toni25} presented such a machine-learning classification, calibrated on the COSMOS15 sample and applied to the COSMOS-Web galaxy catalog (see Sect.~\ref{sec:data}), to study the quiescent galaxy fraction across $0<z<3.7$. 
This selection technique, based on the
\texttt{eXtremeGradientBoosting} classifier \citep[\texttt{XGB},][]{Chen16}, offers a valuable benchmark for the probabilistic classification implemented in the \texttt{PointProbability} class of \texttt{FOMO-Z} (Sect.\ \ref{sec:method:point_prob}), allowing a direct comparison with a machine-learning-based classification of the same galaxy sample. \\
\indent The classification of \citet{Toni25} relies on rest-frame $NUV$, $r$, $J$, and $K$ magnitudes from COSMOS15. To ensure a consistent comparison, we used the same catalog and photometric bands to estimate quiescence probabilities with \texttt{PointProbability}. The classifier of \citet{Toni25} is based on ground-truth labels built from four diagnostics widely used in the literature: the $NUVrJ$ \citep{Ilbert13} and $NUVrK$ \citep{Davidzon16} rest-frame color cuts, a fixed specific star formation rate (sSFR) threshold of $10^{-10.5}$ yr$^{-1}$ \citep{Salim18}, and the Sa-color evolution method \citep{Radovich20}, which tracks the observed color evolution of a modeled Sa-type galaxy across the 4000\,\AA\ break. Galaxies were labeled as quiescent or star-forming according to the majority of these diagnostics, while ambiguous cases were assigned to a third class. This class includes both galaxies with uncertain classifications and potential green-valley objects, and was excluded from the training set of \citet{Toni25}.\\
\indent We adopted exactly the same COSMOS15 reference sample and class labels as \citet{Toni25} for \texttt{FOMO-Z}, retaining only galaxies labeled as star-forming or quiescent and excluding ambiguous cases. We combined the reference and test color uncertainties in quadrature to account for photometric errors in both samples (Eq.~\ref{eq:tolerance}).\footnote{Neither COSMOS15 nor COSMOS-Web provides uncertainties on rest-frame absolute magnitudes. We approximated them using the magnitude error of the observed filter closest in logarithmic wavelength to $(1+z)\lambda_X$, where $\lambda_X$ is the rest-frame wavelength of band $X$. Errors corresponding to $S/N<3$ were replaced by the mean error for $S/N\geq3$ measurements in the same filter and catalog. This approximation includes only photometric errors, neglecting uncertainties in redshift, $K$-corrections, and spectral energy distribution modeling.} This approximation neglects possible correlations between the two catalogs arising from shared photometric data or common systematics. \\
\indent For this test, we defined the target sample rule (Eq.~\ref{eq:R}) as
\begin{equation}
\mathcal{R} : \delta_{\rm quiesc} = 1\,,
\end{equation}
where $\delta_{\rm quiesc}=1$ if the reference galaxy is labeled as quiescent, and $\delta_{\rm quiesc}=0$ otherwise. We selected the most informative colors using the method described in Appendix~\ref{app:features}. Starting from the rest-frame $NUV$, $r$, $J$, and $K$ magnitudes in the COSMOS15 reference sample, this procedure retained only the colors $(NUV-r)$ and $(r-J)$. The other colors are classified as either redundant or non-informative. For each galaxy in the COSMOS-Web test sample, we therefore identified the COSMOS15 reference galaxies whose colors are compatible with those of the test galaxy within the adopted photometric uncertainties, using only these two colors (Eq.~\ref{eq:set_similar}). The resulting set of reference galaxies is then used to estimate the probability that the test galaxy is quiescent (Eq.~\ref{eq:point_probability}). We verified that the two selected colors yield results consistent with those obtained using all available colors, showing that feature selection reduces the number of colors required while preserving the resulting probability estimates. When more bands are available (see, e.g., Sect.\ \ref{sec:results:soms}), feature selection not only reduces computational cost but can also be essential to prevent redundant or uninformative colors from overly restricting the set of compatible reference objects. \\
\indent The comparison with the classification of \citet{Toni25} is shown in Fig.~\ref{fig:Toni26_comparison}. Both methods identify similar regions of color space as predominantly quiescent or star-forming. However, \texttt{FOMO-Z} produces a sharper separation between the two populations, with mean quiescence probabilities closer to zero or unity and a narrower transition region. This behavior reflects the binary labeling of the reference sample, which combines all four diagnostics described above. The broad agreement with the color--color cuts of \citet{Ilbert13}, shown in the figure as white lines, is consistent with their contribution to this combined labeling. Assessing whether the sharper separation corresponds to higher classification accuracy requires an independent test sample with reliably established quiescent and star-forming labels. We verified that the same qualitative comparison holds when the sample is projected onto the $(NUV-r)$--$(r-K)$ plane, indicating that the observed behavior is not specific to the projection onto the feature-selected colors $(NUV-r)$ and $(r-J)$.

\subsection{Comparison with self-organizing maps}\label{sec:results:soms}
Self-organizing maps (SOMs; \citealt{Kohonen1982,Wright25}) are unsupervised machine-learning algorithms that have been widely used in photometric surveys to reconstruct the redshift distributions of galaxy samples. Galaxies are first assigned to tomographic redshift bins based on noisy photo-$z$ estimates, and SOMs are then used to infer the underlying redshift distributions using spectroscopic samples as the ground truth. SOM-based methods are widely adopted in current cosmic shear and galaxy clustering analyses \citep{Heymans21,Wright25_cosmo}.\\
\indent In this section, we compare the redshift distributions reconstructed with \texttt{FOMO-Z} with those obtained using SOMs in \citet{Wright25_cosmo}. We adopted the same tomographic bins and used the KiDZ sample (Sect.\ \ref{sec:data}), which provides both photometric and spectroscopic redshift estimates for each galaxy, as the reference sample, and the KiDS-Legacy shear sample \citep{Wright25_cosmo} as the test sample. We used photometry in the $u$, $g$, $r$, $i$, $Z$, $Y$, $J$, $H$, and $K_{\rm s}$ bands in both samples. As discussed by \citet{Wright25}, the KiDZ photometry retains the measurement uncertainties characteristic of the KiDS cosmological survey.\footnote{The KiDZ and KiDS shear catalogs used here do not include magnitude uncertainties. We retrieved them by matching sources to the KiDS photometric catalog within $1''$ and requiring $|\Delta Z_{\rm B}|\leq0.001$. All KiDS sources had a match. For unmatched KiDZ sources, we imputed uncertainties independently in each band by randomly drawing from the valid measured errors of up to 100 matched sources in the same sample nearest in magnitude.} In this analysis, \texttt{FOMO-Z} uses only colors as input features. The KiDS-Legacy SOMs, instead, use a set of inputs comprising all colors and the total $r$-band magnitude, which is not included in our analysis. Future versions of \texttt{FOMO-Z} will support additional features, such as galaxy magnitudes and sizes. \\
\indent We supplied the reference and test samples to the \texttt{Reconstruction} class, specifying the photometric redshift point estimate \texttt{Z\_B}, corresponding to the peak of the posterior distribution, as the selection variable and the tomographic bin edges as the selection limits. These choices define the selection rule $\mathcal{R}_{\rm obs}$ in Eq.~\eqref{eq:reconstruction_selection}. Both samples were supplied without prior tomographic cuts, allowing the feature-selection procedure described in Appendix~\ref{app:features} to identify colors that distinguish KiDZ galaxies inside and outside each tomographic bin (see Sect.\ \ref{sec:method:reconstruction} for details). Color matching, based on Eq.~\eqref{eq:set_similar}, was then performed between the KiDS and KiDZ subsets selected within the same photometric redshift interval. The spectroscopic redshifts of the matched KiDZ analogs were subsequently used to reconstruct the KiDS redshift distributions according to Eq.~\eqref{eq:reconstruction_counts}.\\
\indent As shown in Eq.~\eqref{eq:reconstruction_weights}, reference prior weights determine the relative contributions of compatible analogs, while test weights determine each galaxy's contribution to the reconstructed population distribution. Following \citet{Wright25}, we assigned each reference galaxy a prior weight equal to the product of its prior-volume and shape-measurement weights. Test galaxies were weighted by their shape-measurement weights. We excluded SOM-derived gold weights from both samples to keep the \texttt{FOMO-Z} reconstruction independent of the SOM analysis. A code snippet reproducing this calibration is reported in Fig.\ \ref{fig:code_som} of Appendix \ref{app:usage}.\\
\indent Figure~\ref{fig:som} compares the redshift distributions reconstructed with \texttt{FOMO-Z} and the SOM-based estimates of \citet{Wright25}. As discussed above, the SOM estimates include gold weighting, which is omitted from our reconstruction. Neither set of distributions includes simulation-based bias corrections. The comparison therefore concerns the estimates before this calibration. The two methods yield qualitatively similar distributions across the six tomographic bins, with closely aligned main peaks. The largest visible discrepancy occurs in the highest-redshift bin, where \texttt{FOMO-Z} assigns more weight to higher redshifts and yields a higher mean redshift than the SOM estimate. These differences reflect a combination of methodological choices, input features, and weighting schemes, including the different weighting of the target population, and cannot be attributed to the reconstruction method alone. Their impact on cosmological constraints from cosmic shear and galaxy clustering will be investigated in future work.\\
\indent This comparison illustrates the use of \texttt{FOMO-Z} for redshift-distribution reconstruction with distinct reference and test samples. The framework allows weights in both samples and selects informative colors for each tomographic bin, providing a flexible alternative approach to population-level inference from photometric data. Beyond the use of spectroscopic reference samples, \texttt{FOMO-Z} can incorporate reference PDZs through their integrated probability in each redshift bin, rather than relying solely on point estimates (Eq.\ \ref{eq:reconstruction_pdz}). This enables the use of photometric reference samples such as COSMOS while propagating their redshift uncertainties.

\section{Conclusions}\label{sec:conclusions}
We presented \texttt{FOMO-Z} (Flexible Optimizer of Multi-dimensional Operators and photo-$z$s), an open-source Python package for galaxy classification, probabilistic inference, and reconstruction of population distributions from photometric observables. The methods use reference samples to relate observable quantities to galaxy properties through explicit, interpretable criteria. The methods support reference prior weights and redshift probability distributions, as well as test-sample weights when reconstructing population distributions. \\
\indent The \texttt{binary\_classification} subpackage implements two complementary approaches that maximize completeness given a prescribed minimum purity. The \texttt{LinearClassification} class calibrates linear color cuts and is particularly suited to weak lensing background selection, as demonstrated in Sect.~\ref{sec:results:background}. The \texttt{GridClassification} class provides more flexible selection regions through discretized color--color grids. In the applications presented here, its role is to support the feature-selection procedure described in Appendix~\ref{app:features}, which identifies informative colors for the probabilistic methods. \\
\indent The \texttt{probability} subpackage extends this framework beyond binary selections. The \texttt{PointProbability} class estimates individual probabilities of satisfying user-defined criteria by identifying color-compatible reference objects within tolerances determined by the photometric uncertainties. We demonstrated its application to distinguishing quiescent and star-forming galaxies in Sect.~\ref{sec:results:quiescent}. The \texttt{ProbabilityFunction} class estimates PDFs of continuous galaxy properties by fitting monotone cumulative probability estimates and differentiating the fitted functions. Its application to photo-$z$ estimation illustrates how the same matching framework can characterize individual galaxies beyond point-estimate probabilities. \\
\indent The \texttt{Reconstruction} class estimates property distributions of selected test populations by transferring test-object weights to compatible reference analogs. Applied to the KiDS shear sample with KiDZ as the reference, this method yields redshift distributions broadly similar to those obtained through SOM-based calibration (Sect.~\ref{sec:results:soms}). Reference PDZs can also contribute to reconstructed redshift distributions, allowing the use of photometric reference samples when their redshift uncertainties are well characterized. \\
\indent These applications illustrate the flexibility of \texttt{FOMO-Z} across galaxy selection and population characterization tasks. Its computational performance is examined in Appendix~\ref{app:performance}. The computational efficiency of the implementations makes them well suited also for optimizing the trade-off between classification completeness and purity, for instance in maximizing the signal-to-noise ratio of weak lensing measurements (Sect.~\ref{sec:results:maximize}). Another key application of this framework is the optimization of the purity--completeness trade-off in flux-limited surveys. \citet{Quai25} showed how tuning the minimum purity threshold to maximize completeness allows the construction of galaxy samples suitable for recovering galaxy-evolution scaling relations, in both the Wide and Deep Euclid surveys \citep{Mellier25}. \\
\indent Regarding background galaxy selections for weak lensing analyses, binary color cuts provide a practical solution for measurement pipelines that operate on preselected source catalogs and do not explicitly incorporate background-membership probabilities. While \texttt{LinearClassification} provides such practical binary selections (see Sects.\ \ref{sec:results:background} and \ref{sec:results:maximize}), a more informative approach is to incorporate the individual background probabilities estimated by \texttt{PointProbability} directly into the source weights, for example through
\begin{equation}\label{eq:new_gt}
    g_{\rm t}(R_j)=
    \frac{\sum_{i\in j} w_i\,p_i\,e_{\rm t,i}}
     {\sum_{i\in j} w_i\,p_i}\,.
\end{equation}
Here, the sums run over sources in radial annulus $j$. The factor $p_i=\Pr(z_i>z_{\rm l})$ is the probability that source $i$ lies behind a lens at redshift $z_{\rm l}$, derived using \texttt{PointProbability}, while $w_i$ is its ellipticity measurement weight, and $e_{\rm t,i}$ is its tangential ellipticity. Shape-calibration corrections, intrinsic alignments, and magnification are omitted for simplicity. Incorporating the $p_i$ probabilities into measurement pipelines would allow sources to contribute according to their estimated background membership, while requiring residual contamination and the source redshift distribution to be modeled consistently with the adopted weights. We suggest the following approach to reconstruct the redshift distribution of the source sample, $n(z_{\rm g}|z_{\rm l})$, when Eq.~\eqref{eq:new_gt} is used as the reduced tangential shear estimator. The $p_i$ probabilities derived with \texttt{PointProbability} can be passed to \texttt{Reconstruction} as test weights, $w^i_{\rm test} = w_i\,p_i$ (see Eq.~\ref{eq:reconstruction_weights}), so that $n(z_{\rm g}|z_{\rm l})$ describes the source population with the same weighting adopted in Eq.~\eqref{eq:new_gt}. This procedure is not circular, as the $p_i$ values only define the weighting of the source sample, whose redshift distribution is then inferred by \texttt{Reconstruction}. The resulting $n(z_{\rm g}|z_{\rm l})$ enters the computation of $\Sigma_{\rm crit}$ in the model (Eq.~\ref{eq:sigma_crit}). In addition, foreground sources dilute the lensing signal, and this contamination must be included in the model (see Sect.~\ref{sec:results:maximize}). The fraction of contaminants in the weighted source sample is
\begin{equation}
f_{\rm c}(z_{\rm l}) = \int_{0}^{z_{\rm l}}{\rm d}z_{\rm g}\, n(z_{\rm g}|z_{\rm l})\,,
\end{equation}
where $n(z_{\rm g}|z_{\rm l})$ is normalized to unity, and the lensing signal is diluted by a factor $1-f_{\rm c}(z_{\rm l})$, corresponding to the purity $\mathcal{P}$ in Eq.~\eqref{eq:g_cen}. This approach assumes that $z_{\rm l}$ is accurately known. If the lens redshift is highly uncertain, $p_i$, $f_{\rm c}$, and $n(z_{\rm g}|z_{\rm l})$ become uncertain as well, since they all depend on $z_{\rm l}$, and their interpretation is no longer straightforward.\\
\indent Some applications require methods complementary to \texttt{FOMO-Z}. In cluster weak lensing analyses, contamination by cluster members may require dedicated red-sequence models or contamination estimates derived from large cluster samples \citep{Medezinski10,Grandis24}. Spatial cross-correlation measurements can also provide complementary constraints on reconstructed redshift distributions \citep{Menard13}. \\
\indent The reliability of the methods presented in this work depends on the coverage and quality of the reference sample. The latter must adequately represent the galaxy populations present in the test sample, and their photometry must be consistently calibrated. The KiDZ sample (Sect.\ \ref{sec:data}) provides a useful example, because it combines survey photometry with spectroscopic redshifts. More generally, reference and test photometry must be measured on a consistent system, and the reference sample must cover the color range occupied by the test sample. Realistic simulations with known galaxy properties can be used to quantify the accuracy of the inferred probabilities and distributions, including biases caused by differences between the reference and test samples. \\
\indent Future developments will extend the input features beyond colors to include observables such as magnitudes, luminosity, and galaxy size. For instance, while the application of \texttt{ProbabilityFunction} presented in Sect.~\ref{sec:method:prob_func} focuses on photometric redshifts, extending its use to stellar masses will require incorporating luminosity information: colors constrain the stellar mass-to-light ratio, but do not determine the luminosity normalization needed to infer stellar mass. Together with the methods presented here, these developments will broaden the applications of \texttt{FOMO-Z} to current and upcoming photometric surveys.

\begin{acknowledgements}
I warmly thank L.\ Moscardini, F.\ Marulli, M. Bolzonella, M. Radovich, C.\ Giocoli, M. Sereno, G. Castignani, M.\ Maturi, and G. Toni for the insightful discussions. I acknowledge the support by the Italian Ministry of University and Research (grant FIS2023-01611, CUP C53C25000300001). I acknowledge the financial contribution from the grant ASI n. 2024-10-HH.0 “Attività scientifiche per la missione Euclid – fase E”.
\end{acknowledgements}

\bibliography{aanda}

\appendix

\section{Installation and examples}\label{app:usage}
\begin{figure}[h]
\centering
\begin{forest}
  for tree={
    font=\ttfamily,
    grow'=0,
    child anchor=west,
    parent anchor=south,
    anchor=west,
    calign=first,
    edge path={
      \noexpand\path [draw, \forestoption{edge}]
      (!u.south west) +(7.5pt,0) |- (.child anchor) \forestoption{edge label};
    },
    before typesetting nodes={
      if n=1
        {insert before={[,phantom]}}
        {}
    },
    fit=band,
    before computing xy={l=15pt},
    s sep=2pt,
    inner ysep=1.5pt
  }
[FOMO-Z/
  [data/]
  [examples/]
  [fomoz/
    [auxiliary/]
    [binary\_classification/]
    [data/]
    [probability/]
  ]
  [tests/]
]
\end{forest}
\caption{Directory structure of the \texttt{FOMO-Z} Python package.}
\label{fig:package-structure}
\end{figure}
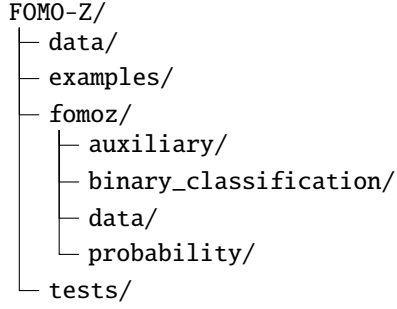

\begin{table}[h]
\centering
\caption{Package requirements.}
\begin{tabular}{|c|c|c|c|}
\hline
\textbf{Package} & \textbf{Version constraint} & \textbf{Base} & \textbf{Development} \\
\hline
numpy & $\geq 1.21.0$ & \checkmark & \checkmark \\
\hline
scipy & $\geq 1.7.0$ & \checkmark & \checkmark \\
\hline
pandas & $\geq 1.3.0$ & \checkmark & \checkmark \\
\hline
astropy & --- & \checkmark & \checkmark \\
\hline
matplotlib & $\geq 3.4.0$ & \checkmark & \checkmark \\
\hline
tqdm & $\geq 4.62.0$ & \checkmark & \checkmark \\
\hline
PyYAML & $\geq 5.4.0$ & \checkmark & \checkmark \\
\hline
numexpr & $\geq 2.7.0$ & \checkmark & \checkmark \\
\hline
numba & $\geq 0.56.4$ & \checkmark & \checkmark \\
\hline
joblib & $\geq 1.3.0$ & \checkmark & \checkmark \\
\hline
threadpoolctl & $\geq 3.1.0$ & \checkmark & \checkmark \\
\hline
pytest & $\geq 6.0.0$ & & \checkmark \\
\hline
black & $=24.8.0$ & & \checkmark \\
\hline
pydocstyle & $\geq 6.0.0$ & & \checkmark \\
\hline
\end{tabular}
\tablefoot{
The first two columns list package names and version constraints.
A dash indicates that no explicit version constraint is imposed.
Checkmarks indicate dependencies of the base and development installations.
The \texttt{requirements.txt} file specifies the packages needed to run
\texttt{FOMO-Z}.
The \texttt{requirements-dev.txt} file includes these base requirements
and adds tools for testing (\texttt{pytest}), code formatting
(\texttt{black}), and docstring style checking (\texttt{pydocstyle}).
}
\label{tab:requirements}
\end{table}

\begin{figure}[t]
\centering
\begin{lstlisting}
import pandas as pd
from fomoz.data import (
    Dataset, SelectionDefinition
)
from fomoz.binary_classification import (
    LinearClassification,
    Classifier, PlotCalibration
)

# Load the reference sample, having the entries
# 'g', 'r', 'i', 'redshift'
ref_sample = pd.read_csv('reference.csv')

# Define the z_l point(s) and photometric bands
z_l = 0.65
bands = ['g', 'r', 'i']

# Dataset and selection criteria
data = Dataset(
    ref_sample, band, z_gal_string="redshift"
)
select = SelectionDefinition(z_lim=z_l)

# Calibration of linear color cuts
P_min = 0.96  # global purity threshold

calib = LinearClassification(
    data, select, P_min
)
calib.process()

# Get the histogram of any property in the
# reference sample, e.g. redshift, after the
# application of the selection mask
bin_edges = np.linspace(0, 2, 101)
calib.get_distribution(0, "redshift", bin_edges)

# Print the total completeness, purity, and
# failure rate (with bootstrap errors)
print("C:", calib.C, "+-", calib.C_std)
print("P:", calib.P, "+-", calib.P_std)
print("F:", calib.F, "+-", calib.F_std)

# Plot the results in color-color spaces.
# It requires a Classifier instance.
color_selection = Classifier()
color_selection.load_parameters(calib)

plotter = PlotCalibration(color_selection, data)
plotter.plot_all_color_distributions(0)

# [Optional] Write calibration parameters
# and metrics to file
calib.write(
    './output/',  # output directory
    'griz_selection'  # output file
)
\end{lstlisting}
\caption{Code snippet showing the usage of the \texttt{binary\_classification} subpackage. It reproduces the selection displayed in Fig.\ \ref{fig:oguri12}, based on the \texttt{LinearClassification} class.}
\label{fig:code_oguri}
\end{figure}

\begin{figure*}[t]
\centering
\begin{lstlisting}
import pandas as pd
import numpy as np
from fomoz.data import Dataset
from fomoz.probability import Reconstruction

# Load both parent samples without tomographic cuts.
# Each magnitude column must have a corresponding <band>_error column.
ref_sample = pd.read_csv('KiDZ.csv')
test_sample = pd.read_csv('KiDS.csv')

bands = ['u', 'g', 'r', 'i', 'Z', 'Y', 'J', 'H', 'K']

# Reference prior weights and test population weights. No gold weights.
ref_weights = (ref_sample['prior_volume_weight'] * ref_sample['shape_weight'])
test_weights = test_sample['shear_weight']

# Dataset objects
reference = Dataset(ref_sample, bands, weights=ref_weights)
test = Dataset(test_sample, bands, weights=test_weights)

# Select both samples by their photometric redshift point-estimates (Z_B).
reconstruction = Reconstruction(
    reference, test, selection_variable='Z_B', selection_min=0.10, selection_max=0.42
)

# Select informative colors using the full reference sample.
reconstruction.select_features_grid()

# Match the selected subsets and transfer test weights to analogs.
reconstruction.process()

# Reconstruct the spectroscopic redshift distribution:
# 100 bins, normalized over the range [0, 2.3].
bin_edges = np.linspace(0.0, 2.3, 101)
n_z = reconstruction.get_distribution('z_spec', bin_edges, normalize=True)
\end{lstlisting}
\caption{
Example of the \texttt{Reconstruction} workflow used in Sect.~\ref{sec:results:soms}, with KiDZ as the reference sample and the KiDS-Legacy shear sample as the test sample. Both parent samples are supplied before applying the tomographic selection $z_{\rm phot}\in(0.10,0.42]$, defined using the photometric redshift point estimate \texttt{Z\_B}. Reference prior weights combine prior-volume and shape-measurement weights, while test weights are the shape-measurement weights. No SOM-derived gold weights are used. The \texttt{select\_features\_grid} method implements the feature selection described in Appendix \ref{app:features}. The output \texttt{n\_z} is a normalized redshift distribution constructed from the reference \texttt{z\_spec} values using the transferred weights of Eq.~\eqref{eq:reconstruction_weights}.}
\label{fig:code_som}
\end{figure*}

\noindent The \texttt{FOMO-Z} package can be installed from source by cloning the public repository available on GitHub\footnote{\url{https://github.com/GiorgioLesci/FOMO-Z}}:
{\small\begin{verbatim}
git clone https://github.com/GiorgioLesci/FOMO-Z.git
\end{verbatim}}
\noindent The package, whose structure is illustrated in Fig.\ \ref{fig:package-structure}, can then be installed using \texttt{pip}:
{\small\begin{verbatim}
cd FOMO-Z
pip install .
\end{verbatim}}
\noindent Package requirements are listed in Table \ref{tab:requirements}. Although installation in the system environment is possible, we recommend using an isolated environment to avoid dependency conflicts. For example, a \texttt{conda} environment can be created as follows:
{\small\begin{verbatim}
conda create -n fomoz python=3.9 "numpy<2" scipy \
    astropy pyyaml -c conda-forge
conda activate fomoz
pip install .
\end{verbatim}}
\noindent This environment can be activated at any time with:
{\small\begin{verbatim}
conda activate fomoz
\end{verbatim}}
\noindent Figures~\ref{fig:code_oguri} and \ref{fig:code_som} provide example code snippets illustrating how to reproduce some of the results presented in Sect.~\ref{sec:results}.

\section{RAM usage and execution time}\label{app:performance}
\begin{figure*}[t]
\centering\includegraphics[width = \hsize, height = 8.5cm] {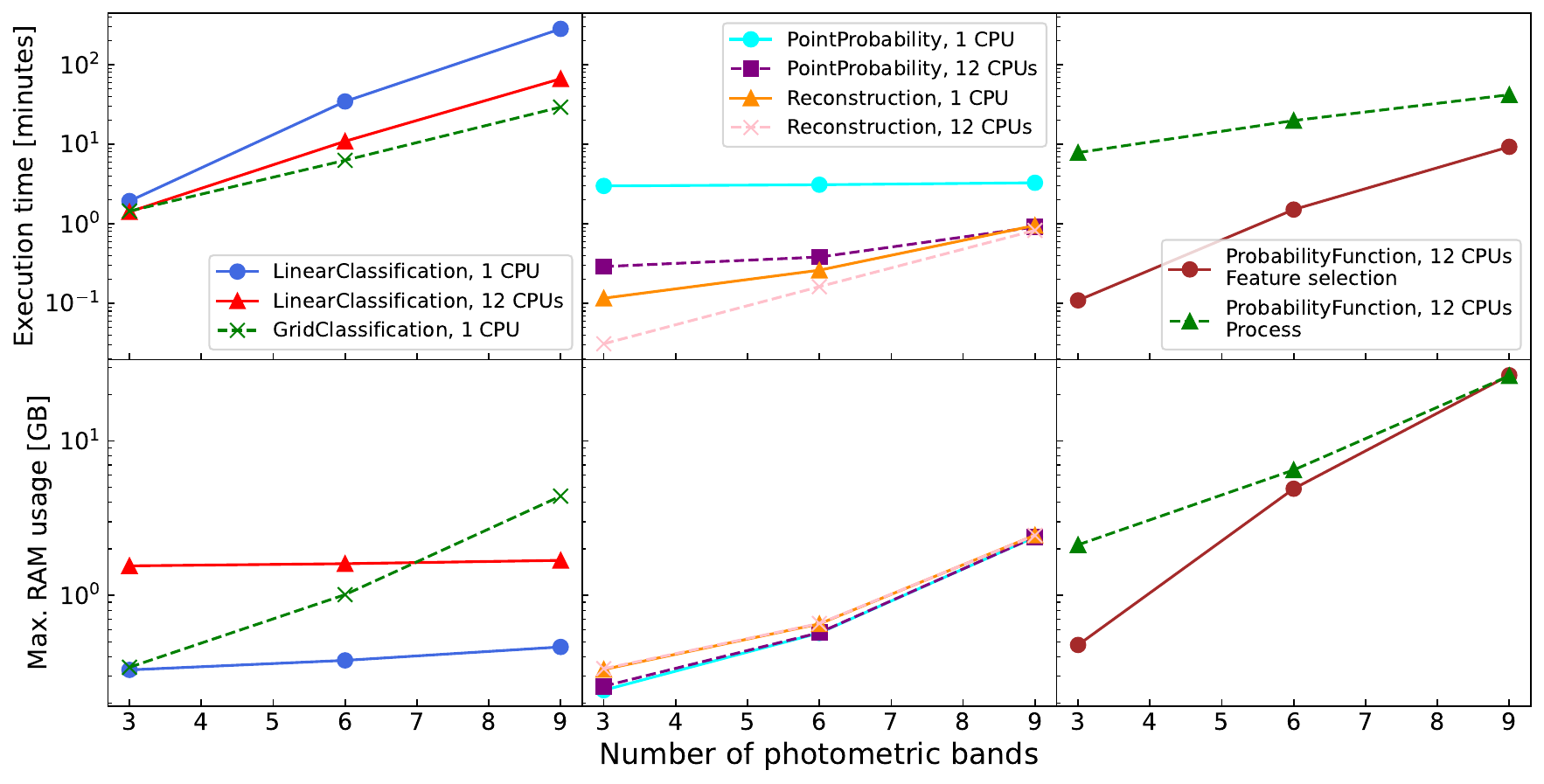}
\caption{Execution time (top) and peak memory usage (bottom) of the FOMO-Z routines as a function of the number of photometric bands, using the COSMOS20 catalog as the reference sample. The left panels show results for \texttt{LinearClassification} using 1 and 12 CPUs and for \texttt{GridClassification}, which is not parallelized. These binary-classification calibrations need only be performed once for a given configuration and can subsequently be applied to other samples through the \texttt{Classifier} class. The middle panels show results for \texttt{PointProbability} and \texttt{Reconstruction} using 1 and 12 CPUs. The right panels show the feature-selection and processing stages of \texttt{ProbabilityFunction}, measured separately using 12 CPUs. In the middle and right panels, the COSMOS20 catalog is used as both the reference and test sample. Colors, markers, and line styles identify the routines and CPU configurations as indicated in the legends.}
\label{fig:performance}
\end{figure*}
Figure \ref{fig:performance} reports the execution time and peak memory usage of the implementations presented in this work, as a function of the number of photometric bands. The COSMOS20 catalog (Sect.\ \ref{sec:data}) is used as the reference sample for all routines and also as the test sample for \texttt{PointProbability}, \texttt{Reconstruction}, and \texttt{ProbabilityFunction}. Unless otherwise specified, we use the default \texttt{FOMO-Z} parameters. \\
\indent The \texttt{LinearClassification} processing is parallelized via Python's \texttt{multiprocessing} module, which distributes the evaluation of color cut parameter values across multiple CPU cores. As evident from the left panels of Fig.\ \ref{fig:performance}, parallelization is recommended if the number of photometric bands used for the calibration is $\gtrsim3$. If 12 CPUs are used, the execution time is about 2 minutes, 15 minutes, and 90 minutes for 3, 6, and 9 photometric bands, respectively. \texttt{GridClassification}, instead, relies on vectorized array operations without explicitly distributing the computation across CPU cores. Nevertheless, it is faster than \texttt{LinearClassification} even when the latter uses 12 CPUs. The execution time of both methods increases with the number of photometric bands, as more color combinations must be evaluated. \\
\indent The processing stages of \texttt{PointProbability} and \texttt{Reconstruction}, illustrated in the middle panels of Fig.\ \ref{fig:performance}, are parallelized over blocks of test objects using threads, reducing the execution time when 12 CPUs are used. Both methods perform feature selection once for the specified target selection. Its contribution is small for fewer bands but becomes dominant for \texttt{Reconstruction} at nine bands, causing the execution times obtained with 1 and 12 CPUs to converge. In contrast, \texttt{ProbabilityFunction} requires feature selection at each sampled threshold of the random variable, making this stage a relatively substantial computational contribution. Its feature selection and processing times are therefore shown separately in the right panels of Fig.\ \ref{fig:performance}, both using 12 CPUs. Feature selection is parallelized across thresholds, while processing includes the estimation of cumulative probabilities followed by spline fitting (Eq.~\ref{eq:cdf_spline}), both parallelized over blocks of test objects using threads and processes, respectively.\\
\indent The RAM usage is very low for all implementations. \texttt{LinearClassification} keeps RAM usage almost constant as a function of the number of photometric bands. This is because single color conditions in Eq.\ \eqref{eq:all_cond} that yield a purity well below the requested threshold are not stored in memory: only the color cuts close to the optimal ones are kept. This drastically reduces unnecessary memory usage. The other classes, instead, show a remarkable increase in memory usage with the number of bands. However, RAM usage remains relatively modest in all cases, ranging from around 1 GB with 3 bands to approximately 10 GB with 9 bands. The reported RAM usage excludes the memory occupied by the reference and test samples. \\
\indent We remark that the \texttt{LinearClassification} and \texttt{GridClassification} calibrations are computed only once, and are then applied to any test dataset using the \texttt{Classifier} class. Since this class relies on vectorized array operations over precomputed color conditions and grids, its execution time is negligible compared to that of the calibration step itself.

\section{Color domains and sampling}\label{app:sampling}
In \texttt{FOMO-Z}, the explored domains of the $c_1\dots c_8$ parameters in Eq.\ \eqref{eq:all_cond}, along with the grid domains considered in the \texttt{GridClassification} class (Sect.\ \ref{sec:method:regions}), are empirically determined. For color conditions in Eq.\ \eqref{eq:all_cond} that involve only one color, we consider the following ranges:
\begin{align}\label{eq:linear_ranges}
c_1,c_2&\in\big[Q_{0.1\%}(x),Q_{99.9\%}(x)\big]\,,\nonumber\\
c_3,c_4&\in[Q_{0.1\%}(y),Q_{99.9\%}(y)]\,,
\end{align}
where $x$ and $y$ are the two different colors in the reference galaxy sample, while $Q_{q\%}(\cdot)$ denotes the empirical $q$-th percentile. By default, \texttt{FOMO-Z} assumes a precision of $\delta c_i=0.05$, which is smaller than the typical magnitude error in COSMOS20. This allows to explore a sufficiently detailed parameter space without missing optimal solutions. \\
\indent The color ranges used to define the grids in \texttt{GridClassification} (Sect.\ \ref{sec:method:regions}) are defined as follows:
\begin{align}
x&\in\big[Q_{0.01\%}(x),Q_{99.99\%}(x)\big]\,,\nonumber\\
y&\in[Q_{0.01\%}(y),Q_{99.99\%}(y)]\,.
\end{align}
Here, the color ranges are wider compared to Eq.\ \eqref{eq:linear_ranges}, to ensure that edge points in the color--color distributions are enclosed within the grid boundaries. By default, the grid precision amounts to 0.15. \\
\indent For the conditions in Eq.\ \eqref{eq:all_cond} involving a linear relation between colors, namely $x \gtrless s_j\,y + c_i$,
the admissible domain depends on both colors and on the slope $s_q$. Defining
\begin{equation}
X(s_j) = x - s_j\,y\,,
\end{equation}
the valid range of $c_p$ (with $i=5,6,7,8$) is defined as follows:
\begin{equation}
c_i \in \big[ Q_{0.1\%}(X(s_j)), Q_{99.9\%}(X(s_j)) \big]\,.
\end{equation}
The precision of the $c_p$ sampling and the set of $s_q$ values are inputs of the \texttt{LinearClassification} class (Sect.\ \ref{sec:method:linear}).

\section{Interpolation of linear color conditions}\label{app:interp}
\begin{figure}[t]
\centering\includegraphics[width = \hsize-1.5cm, height = 4.5cm] {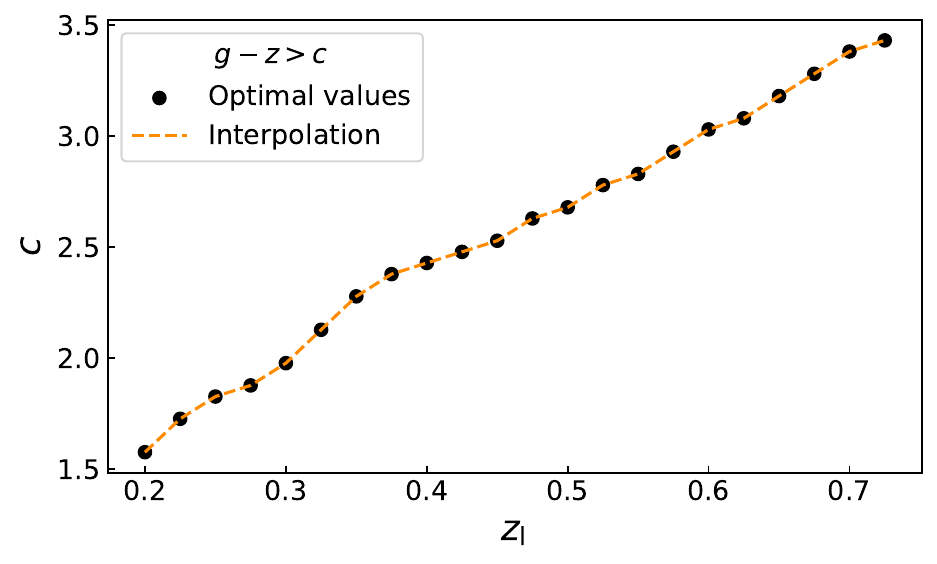}
\caption{Optimal values (black dots) and interpolation (dashed orange curve) of the $c$ parameter from the $g-z>c$ optimal condition, as a function of the lower redshift limit $z_{\rm l}$. The interpolation assumes a spline degree $\kappa_{\rm in}=1$ (Eq.\ \ref{eq:kappa}).}
\label{fig:par_evolution}
\end{figure}

\begin{figure}[t!]
\centering\includegraphics[width = \hsize-1.5cm, height = 7.5cm] {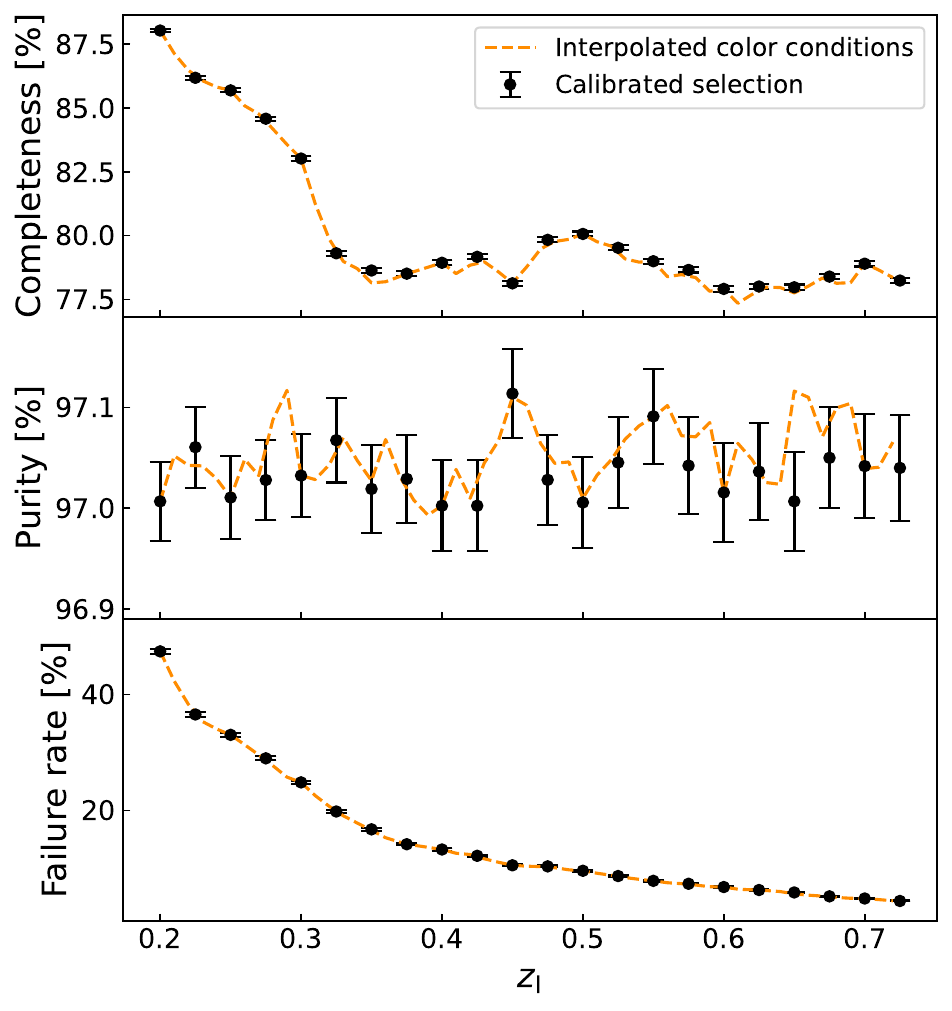}
\caption{Performance metrics of a background selection based on the methods detailed in Sect.\ \ref{sec:method:linear}, derived using COSMOS20 HSC $griz$ photometry and assuming a purity threshold of $\mathcal{P}_{\rm min}=0.97$. The calibration was performed for a set of lower redshift limits spanning $z_{\rm l}\in[0.2,0.75]$, with a precision of $\delta z_{\rm l}=0.025$. The black points represent the recovered metrics, with error bars indicating the bootstrap uncertainties. The dashed orange curves represent the selection obtained by interpolating the color condition parameters as a function of $z_{\rm l}$. The interpolated selection was evaluated assuming a precision of $\delta z_{\rm l}=0.01$.}
\label{fig:CPF}
\end{figure}

If multiple $z_{\rm l}$ points are used in the \texttt{LinearClassification} class (Sect.\ \ref{sec:method:linear}), the parameters $c_p$ and $s_q$ of the optimal color conditions (Eq.\ \ref{eq:all_cond}) are interpolated as a function of $z_{\rm l}$,
using a spline degree, $\kappa$, of
\begin{equation}\label{eq:kappa}
\kappa = \min(\kappa_{\rm in}, n-1)\,.
\end{equation}
Here, $\kappa_{\rm in}$ is the user-specified spline degree and $n$ is the number of $z_{\rm l}$ points. No interpolation is performed when $n\leq1$. A value of $\kappa_{\rm in}=1$ is recommended, as it typically provides sufficient flexibility to capture the underlying trends while avoiding overfitting. As an example, Fig.~\ref{fig:par_evolution} shows the redshift evolution of the parameter $c$ for an optimal color condition, together with its interpolation. In Fig.~\ref{fig:CPF}, we display the performance metrics of an optimal selection calibrated at different $z_{\rm l}$ values, along with the performance obtained by interpolating the color cut parameters between these values. We remark that interpolations must be applied cautiously, as they add numerical noise to purity and completeness estimates, which may be problematic for precision analyses. The bootstrap error estimates described in Sect.\ \ref{sec:method:linear} do not account for such numerical noise.

\section{Improvements over L24}\label{app:L24}
As mentioned in Sect.\ \ref{sec:method:linear}, a first version of the linear color cut methods implemented in the \texttt{LinearClassification} class was introduced by \citetalias{Lesci24}. The main improvement concerns the iterative purity threshold: while \citetalias{Lesci24} assigned the same minimum purity to each individual color condition at each iteration step, here we evaluate the global purity resulting from the combination of all color conditions in Eq.~\eqref{eq:all_cond}. The two approaches yield similar completeness for a given global purity, but the new procedure provides greater control over the final selection, as the old method did not always reach the requested global purity threshold. Furthermore, we extend the method by generalizing the selection criteria to arbitrary galaxy properties beyond redshift, and by allowing reference galaxies to be weighted using prior weights and PDZs (Sect.~\ref{sec:method:sel}). The \texttt{FOMO-Z} implementation also provides the user with explicit control over the color combinations used for the calibration (Sect.~\ref{sec:method:dataset}). Finally, bootstrap uncertainties can be directly estimated through the built-in procedure described in Sect.\ \ref{sec:method:linear}.

\section{Definition of significant bins}\label{app:significance}
As detailed in Sect.\ \ref{sec:method:regions}, in the grid optimization method implemented in \texttt{GridClassification} we compute the probability of finding a galaxy satisfying the set of selection criteria $\mathcal{R}$, denoted as $p_{\rm sel}$ (Eq.\ \ref{eq:p_selection}), in bins of $x$ and $y$ colors. We approximate the uncertainty on $p_{\rm sel}$ using a Binomial model, which is expressed as follows:
\begin{equation}\label{eq:binomial}
P(k \mid n,p) \propto p^k (1-p)^{n-k}\,.
\end{equation}
Here, $k$ is the number of successes, $n$ is the number of trials, and $p$ is the success probability, which we identify as $p \equiv p_{\rm sel}$. For simplicity, in Eq.\ \eqref{eq:binomial} we omit the dependence of $k$, $n$, and $p$ on the color bins. The number of trials is expressed as the Kish effective sample size:
\begin{equation}
n = \frac{\left[\sum_j w_{\rm prior}^j\,\mathcal{S}^j(\delta x,\delta y)\right]^2}{\sum_j \left[w_{\rm prior}^j\,\mathcal{S}^j(\delta x,\delta y)\right]^2}\,.
\end{equation}
The quantities appearing in this equation are detailed in Eq.\ \eqref{eq:p_selection}. Furthermore, the number of successes is expressed as $k = np$. Although Eq.\ \eqref{eq:binomial} formally corresponds to the Binomial likelihood for integer counts, we extend it to the effective quantities $n$ and $k$, and use it as an approximation to describe the uncertainty on $p_{\rm sel}$. \\
\indent To assess the statistical significance of the measured values of $p$, we adopt a Bayesian approach. Using Bayes' theorem, the posterior probability distribution of the success probability $p$ is given by
\begin{equation}\label{eq:beta}
P(p \mid k,n) = \frac{P(k \mid p,n)\,P(p)}{P(k \mid n)}\,,
\end{equation}
where $P(k \mid p,n)$ is the Binomial likelihood (Eq.\ \ref{eq:binomial}), $P(p)$ is the prior probability distribution of $p$, and $P(k \mid n)$ is the evidence, which acts as a normalization constant. We assume a uniform prior over the interval $[0,1]$, reflecting the absence of prior information on the success probability\footnote{The number of trials, $n$, does not provide a priori information on the success probability $p$, so $P(p \mid n) \equiv P(p)$.}. Under these assumptions, the posterior distribution in Eq.\ \eqref{eq:beta} is a Beta distribution with parameters $(k+1,n-k+1).$ \\
\indent We define significant bins based on the minimum selection probability threshold, $p_{\rm min}$, attained at the end of the iterative procedure described in Sect.\ \ref{sec:method:regions}. In particular, a bin is considered significant if the following condition is satisfied:
\begin{equation}
\int_0^{p_{\rm min}} \mathrm{d}p\, P(p \mid k,n) < \alpha\,.
\end{equation}
Here, the quantity on the left-hand side represents the posterior probability that the true selection probability of a bin is below the required threshold $p_{\rm min}$. We retain only bins for which this probability is smaller than the parameter $\alpha$, which can be set by the user. Bins failing this condition are excluded from the final selection, after which the masks are updated and the selection metrics are recomputed.

\section{Feature selection}\label{app:features}

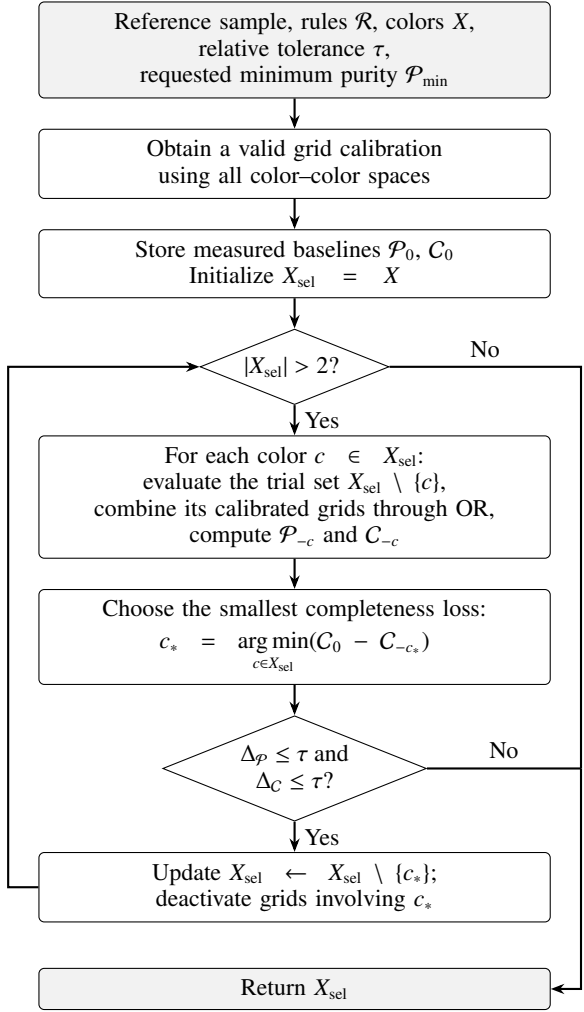
\begin{figure}[t]
    \centering
    \begin{tikzpicture}[
        node distance=4mm,
        font=\footnotesize,
        process/.style={
            rectangle, draw, rounded corners=2pt, align=center,
            text width=0.72\columnwidth, inner sep=4pt
        },
        decision/.style={
            diamond, draw, aspect=2.5, align=center, inner sep=2pt
        },
        terminal/.style={process, fill=gray!10},
        arrow/.style={-{Stealth[length=1.8mm]}, thick}
    ]
    \node[terminal] (input) {
        Reference sample, rules $\mathcal R$, colors $X$,\\
        relative tolerance $\tau$,\\
        requested minimum purity $\mathcal P_{\min}$
    };
    \node[process, below=of input] (calibration) {
        Obtain a valid grid calibration\\
        using all color--color spaces
    };
    \node[process, below=of calibration] (baseline) {
        Store measured baselines $\mathcal P_0$, $\mathcal C_0$\\
        Initialize $X_{\mathrm{sel}}=X$
    };
    \node[decision, below=of baseline] (number) {
        $|X_{\mathrm{sel}}|>2$?
    };
    \node[process, below=of number] (trials) {
        For each color $c\in X_{\mathrm{sel}}$:\\
        evaluate the trial set $X_{\mathrm{sel}}\setminus\{c\}$,\\
        combine its calibrated grids through OR,\\
        compute $\mathcal P_{-c}$ and $\mathcal C_{-c}$
    };
    \node[process, below=of trials] (candidate) {
        Choose the smallest completeness loss:\\[2pt]
        $\displaystyle c_* = \operatorname*{arg\,min}_{c\in X_{\mathrm{sel}}}
        (\mathcal C_0-\mathcal C_{-c_*})$
    };
    \node[decision, below=of candidate] (accept) {
        $\Delta_{\mathcal P}\leq\tau$ and\\
        $\Delta_{\mathcal C}\leq\tau$?
    };
    \node[process, below=of accept] (remove) {
        Update $X_{\mathrm{sel}}\leftarrow X_{\mathrm{sel}}\setminus\{c_*\}$;\\
        deactivate grids involving $c_*$
    };
    \node[terminal, below=6mm of remove] (output) {
        Return $X_{\mathrm{sel}}$
    };

    \draw[arrow] (input) -- (calibration);
    \draw[arrow] (calibration) -- (baseline);
    \draw[arrow] (baseline) -- (number);
    \draw[arrow] (number) -- node[right] {Yes} (trials);
    \draw[arrow] (trials) -- (candidate);
    \draw[arrow] (candidate) -- (accept);
    \draw[arrow] (accept) -- node[right] {Yes} (remove);

    \coordinate[left=4mm of trials.west] (loop);
    \draw[arrow] (remove.west)
        -- (remove.west -| loop)
        |- (number.west);

    \coordinate[right=4mm of trials.east] (exit);
    \coordinate (join) at (accept.east -| exit);
    \draw[thick] (number.east)
        -- node[above] {No} (number.east -| exit)
        -- (join);
    \draw[thick] (accept.east)
        -- node[above] {No} (join);
    \draw[arrow] (join)
        |- (output.east);
    \end{tikzpicture}
    \caption{
        Feature selection algorithm. The input minimum purity
        $\mathcal P_{\min}$ (default $0.9$) is distinct from the measured
        initial purity $\mathcal P_0$. The trial metrics $\mathcal P_{-c}$ and $\mathcal C_{-c}$ are evaluated after
        removing one candidate color from the current retained set $X_{\rm sel}$: previously removed colors remain excluded. Relative changes
        $\Delta_{\mathcal P}=|\mathcal P_{-c_*}-\mathcal P_0|/\mathcal P_0$
        and $\Delta_{\mathcal C}=|\mathcal C_{-c_*}-\mathcal C_0|/\mathcal C_0$
        are always measured against the fixed initial baselines
        $\mathcal P_0$ and $\mathcal C_0$, with default tolerance $\tau=0.1$.
        Only the candidate color yielding the smallest completeness loss
        is tested for acceptance.
    }
    \label{fig:feature_selection}
\end{figure}

\begin{figure}[t!]
\centering\includegraphics[width = \hsize, height = 6.7cm] {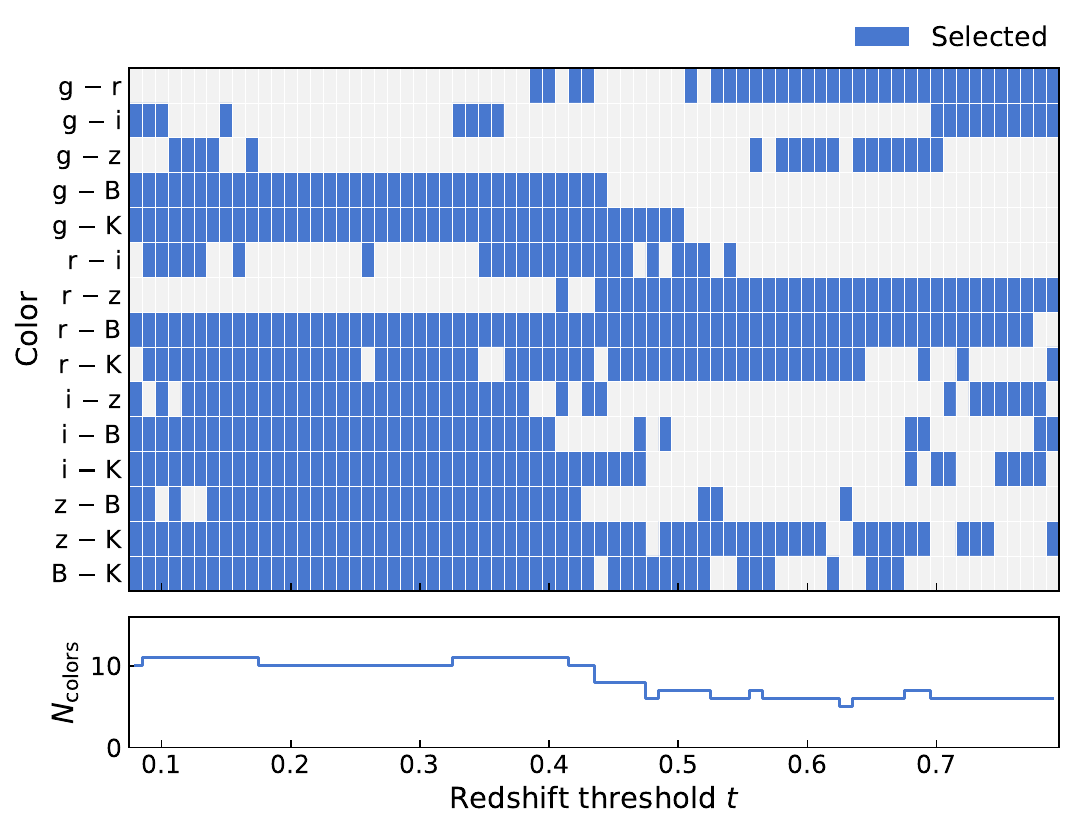}
\caption{Colors retained by the feature-selection procedure in \texttt{ProbabilityFunction} as a function of the redshift threshold $t$, using the $g,r,i,z,B,K$ bands in COSMOS20. Top: blue cells indicate retained colors at each threshold. Bottom: number of retained colors out of the 15 available combinations. The selections are calibrated on the reference sample and apply to all test objects. The map shows which colors are retained, rather than a ranking of their individual importance.}
\label{fig:features}
\end{figure}

The binary classification methods implemented in \texttt{LinearClassification}
(Sect.~\ref{sec:method:linear}) and \texttt{GridClassification}
(Sect.~\ref{sec:method:regions}) do not require a feature-selection step.
Selections calibrated in individual color--color spaces are combined through
a logical `or': an object is selected if it satisfies the selection in at
least one space. Adding a space to this combination therefore does not
impose another matching constraint on objects already selected elsewhere.
In contrast, \texttt{PointProbability}, \texttt{ProbabilityFunction}, and
\texttt{Reconstruction} require reference objects to satisfy the matching
constraints simultaneously in every selected color usable for a given test
object (Eq.~\ref{eq:set_similar}). A color providing little additional
information about target membership can therefore reduce the number of
compatible reference objects without improving discrimination. Feature
selection aims to retain informative colors while limiting unnecessary
restrictions on the matching set. \\
\indent \texttt{FOMO-Z} implements a greedy backward-elimination procedure based on
\texttt{GridClassification}, illustrated in Fig.~\ref{fig:feature_selection}.
For selection rules $\mathcal R$ and the full color set $X$, the procedure
first calibrates a binary selection using all available color--color spaces.
The user supplies a minimum global purity $\mathcal P_{\min}$, equal to
$0.9$ by default. This is a calibration requirement, rather than the exact
purity of the resulting selection. After a successful \texttt{GridClassification} calibration, we
store its measured purity $\mathcal P_0$ and completeness $\mathcal C_0$.
In particular, $\mathcal P_0\geq\mathcal P_{\min}$. Cases where $\mathcal P_0<\mathcal P_{\min}$ are discussed below.
These metrics include the reference prior weights and target memberships
as defined in Sect.~\ref{sec:method:regions}. \\
\indent We initialize $X_{\mathrm{sel}}=X$. At each iteration, the trial removal of
a color $c\in X_{\mathrm{sel}}$ excludes all currently active color--color
grids involving $c$. The remaining calibrated masks are combined through
logical OR, yielding trial purity $\mathcal P_{-c_*}$ and completeness
$\mathcal C_{-c_*}$. Here, $\mathcal P_{-c_*}$ and $\mathcal C_{-c_*}$ are evaluated using the current retained set $X_{\mathrm{sel}}\setminus\{c\}$, so all colors removed in previous iterations remain excluded. The masks themselves are not recalibrated during this
elimination step. We choose
\begin{equation}\label{eq:feature_candidate}
    c_* = \operatorname*{arg\,min}_{c\in X_{\mathrm{sel}}}
    \left(\mathcal C_0-\mathcal C_{-c_*}\right),
\end{equation}
that is, we select the color $c_*$ whose removal produces the smallest loss in completeness relative to the fixed initial baseline $\mathcal C_0$, or equivalently, leaves the highest completeness among all candidate removals. Then, we accept its removal only if
\begin{equation}\label{eq:feature_tolerance}
    \Delta_{\mathcal P}\equiv
    \frac{|\mathcal P_{-c_*}-\mathcal P_0|}{\mathcal P_0}\leq\tau,
    \qquad
    \Delta_{\mathcal C}\equiv
    \frac{|\mathcal C_{-c_*}-\mathcal C_0|}{\mathcal C_0}\leq\tau,
\end{equation}
where $\tau$ is the user-specified relative tolerance, equal to $0.1$ by
default. The baseline values $P_0$ and $C_0$ remain fixed throughout the procedure. The absolute differences in
Eq.~\eqref{eq:feature_tolerance} constrain both increases and decreases in
purity, whereas completeness cannot increase when masks are removed from
the OR combination. The procedure stops when the chosen candidate color fails either
condition or only two colors remain. It does not test alternative candidates
after a rejection. If no removal is accepted, all colors are retained. \\
\indent The retained colors define $X_{\mathrm{sel}}$, used by
\texttt{PointProbability} for individual probability estimation and by
\texttt{Reconstruction} for matching reference and test objects when
reconstructing population distributions. In the latter case, feature
selection targets the observed selection interval, rather than the bins
of the reconstructed distribution
(Sect.~\ref{sec:method:reconstruction}). For \texttt{ProbabilityFunction}
(Sect.~\ref{sec:method:prob_func}), feature selection is performed separately
at every sampled threshold $t$, allowing $X_{\mathrm{sel}}(t)$ to vary
across the reconstructed property range. \\
\indent In \texttt{ProbabilityFunction}, we evaluate at each sampled threshold $t$ the prior-weighted fraction of reference objects satisfying the cumulative criterion, $\mathcal R_t:x<t$, as follows:
\begin{equation}\label{eq:feature_reference_fraction}
    \eta(t)=
    \frac{\sum_j w_{\rm sel}^j(\mathcal R_t)}
         {\sum_j w_{\rm prior}^j}.
\end{equation}
The selection weights are defined as in Eq.~\eqref{eq:w_selection}. For redshift estimation, they incorporate the integrated reference PDZs when available. Otherwise, including for all other reconstructed properties, they reduce to $w_{\rm sel}^j(\mathcal R_t)=w_{\rm prior}^j\,\delta^j(\mathcal R_t)$, where $\delta^j(\mathcal R_t)=1$ if $x_j<t$ and $0$ otherwise. Feature selection uses the criterion
\begin{equation}\label{eq:feature_complement}
    \mathcal R_t^{\rm FS}:
    \begin{cases}
        x<t, & \eta(t)\leq 0.5,\\
        x\geq t, & \eta(t)>0.5.
    \end{cases}
\end{equation}
Thus, calibration targets the less abundant class. The transition corresponds to the median of the prior-weighted reference distribution. For the complementary criterion in Eq.\ \eqref{eq:feature_complement}, the selection weights are $w_{\rm prior}^j-w_{\rm sel}^j(\mathcal R_t)$. This choice affects only the retained colors $X_{\mathrm{sel}}(t)$: the subsequent PDF estimation (Sect.\ \ref{sec:method:prob_func}) always uses the original criterion $\mathcal R_t$. \\
\indent All sampled thresholds start from the same input $\mathcal P_{\min}$. If calibration stops without accepting a non-empty selection at a threshold $t$, \texttt{ProbabilityFunction} adopts the colors from the nearest sampled threshold with successful feature selection. If no threshold succeeds, the procedure raises an error. Figure~\ref{fig:features} illustrates how the colors retained by \texttt{ProbabilityFunction} vary with the redshift threshold when reconstructing individual galaxy redshift PDFs. \\
\indent The procedure described here uses the purity and completeness of calibrated binary selections as proxies for feature relevance, providing an interpretable connection between color selection and probabilistic inference. Its greedy approach avoids an exhaustive search over color combinations while retaining a subset that preserves these selection metrics within the adopted tolerance. Future work will assess whether alternative feature-selection strategies offer further improvements in computational efficiency and in the accuracy and calibration of the resulting probability estimates.

\end{document}